\documentclass[seceq]{ptptex}
\usepackage{wrapft}
\usepackage{graphicx}

\newcommand{\ket}[1]{| {#1} \rangle}     
\newcommand{\kket}[1]{| {#1} \rangle\!\rangle}     
\newcommand{\wtilde}[1]{\widetilde{#1}} 

\def\beq{\begin{eqnarray}}
\def\eeq{\end{eqnarray}}
\def\bsub{\begin{subequations}}
\def\esub{\end{subequations}}
\def\b{\begin{equation}}
\def\bs{\begin{split}}
\def\es{\end{split}}
\def\e{\end{equation}}

\begin{document}

\title{A possible framework of the Lipkin model obeying the $su(n)$-algebra in arbitrary 
fermion number. I
}
\subtitle{The $su(2)$-algebras extended from the conventional fermion-pair and determination of the minimum weight states}

\author{Yasuhiko {\sc Tsue}$^{1,2}$, {Constan\c{c}a {\sc Provid\^encia}}$^{1}$, {Jo\~ao da {\sc Provid\^encia}}$^{1}$ and 
{Masatoshi {\sc Yamamura}}$^{1,3}$
}

\inst{$^{1}${CFisUC, Departamento de F\'{i}sica, Universidade de Coimbra, 3004-516 Coimbra, 
Portugal}\\
$^{2}${Physics Division, Faculty of Science, Kochi University, Kochi 780-8520, Japan}\\
$^{3}${Department of Pure and Applied Physics, 
Faculty of Engineering Science, Kansai University, Suita 564-8680, Japan}
}

\abst{
The minimum weight states of the Lipkin model consisting of $n$ single-particle levels and obeying the $su(n)$-algebra 
are investigated systematically. 
The basic idea is to use the $su(2)$-algebra which is independent of the $su(n)$-algebra. 
This idea has been already presented by the present authors in the case of the 
conventional Lipkin model consisting of two single-particle levels and obeying the $su(2)$-algebra. 
If following this idea, the minimum weight states are determined for any fermion number occupying appropriately $n$ single-particle 
levels. 
Naturally, the conventional minimum weight state is included: 
all fermions occupy energetically the lowest single-particle level in the absence of interaction. 
The cases $n=2$, 3, 4 and 5 are discussed in rather detail. 
}



\maketitle

\section{Introduction}

In 1965, at the early stage of the studies of nuclear many-body theories, Lipkin, Meshkov and Glick proposed a schematic model 
for understanding of microscopic structure of nuclear collective vibration \cite{1}. 
Hereafter, we will call it the Lipkin model. 
Naturally, it was an up-to-date problem in those days. 
The Lipkin model treats many-fermion system consisting of two single-particle levels with the same degeneracy as 
each other. 
In this paper, the degeneracy is denoted as $2\Omega$, which is positive even number. 
For this model, we can construct the $su(2)$-algebra in terms of certain bilinear forms 
in single-particle fermion operators under the condition that the total fermion number operator commutes with the $su(2)$-generators. 
The Hamiltonian adopted in this model is expressed as a function of these $su(2)$-generators. 
Concerning total fermion number $N$, the simplest case 
may be the following: 
In the absence of interaction, all fermions fully occupy energetically lower single-particle level, i.e., $N=2\Omega$. 
Following the review article by Klein and Marshalek \cite{KM}, we call this case ``closed-shell" system.
Conventionally, only this case has been investigated. 
With the aid of this model, we are able to obtain a schematic understanding of collective vibrational states of the ``closed-shell" system 
in terms of superposition of particle-hole pair excitations. 
In this case, it is easy to define the particle and the hole operators.

As a natural generalization of the Lipkin model, first, Li, Klein and Dreizler \cite{2} and Meshkov \cite{3} investigated the model consisting of three 
single-particle levels. 
Needless to say, this model is treated in the frame of the $su(3)$-algebra. 
Further, the generalization to the case of $n$ single-particle levels was performed 
mainly by Okubo \cite{4} and Klein \cite{5}. 
The degeneracy of each level is also equal to $2\Omega$. 
Mathematical framework in this case is given by the $su(n)$-algebra 
with the condition that the total fermion number operator commutes with the $su(n)$-generators. 
Needless to say, Hamiltonian should be expressed as a function of the $su(n)$-generators. 
Hereafter, we will call it the $su(n)$-Lipkin model. 
Including the case $n \geq 3$, also only the ``closed-shell" system, i.e., $N=2\Omega$ has been investigated.

We guess that there exist two reasons why only the case $N=2\Omega$ has been investigated. 
One of the reasons may be the following: 
The Lipkin model aims at describing the particle-hole pair type collective vibration and its ideal form may be expected to 
be realized in this case. 
If excessively speaking, any case except the ``closed-shell" system may be not necessary to investigate. 
The second is related to the minimum weight state. 
The Lipkin model is a kind of the algebraic model. 
Therefore, in order to 
complete the description of the model, 
the first task is to determine the minimum weight states. 
The ``closed-shell" system corresponds to the simplest minimum weight state which enables us to formulate various results 
of the Lipkin model quite easily. 
However, in the case of the $su(2)$-Lipkin model, recently, the present authors proposed an idea \cite{6-1}. 
Under this idea, the minimum weight states can be determined in the concrete form 
for the case of any fermion number. 
The prototype of new boson realization of the $su(2)$-algebra in the Lipkin model used in \citen{6-1} 
can be found in \citen{6-2}. 
This idea suggests us that 
we may know the concrete forms of the 
minimum weight states of the $su(n)$-Lipkin model for any fermion number. 
This problem will be discussed in this paper (I). 
However, even if the minimum weight state can be determined, we have still a problem to be solved. 
In the $su(2)$-Lipkin model, the orthogonal set built on a chosen minimum weight state can be easily obtained by operating 
the raising operator successively on the minimum weight state. 
In the case of the $su(n)$-Lipkin model, formally, there exist too many generators which play a role similar to 
that of the raising operator in the $su(2)$-Lipkin model. 
Therefore, in order to make the $su(n)$-Lipkin model workable, we must present any idea for the operators, the role of which 
is similar to that of the $su(2)$-Lipkin model, i.e., the raising operator. 
This problem will be discussed in next paper (II).

Main aim of this paper is to present concrete forms of the 
minimum weight states for any fermion number in the $su(n)$-Lipkin model 
including the ``closed-shell" system. 
Preliminary argument was performed in the recent paper
by the present authors for the $su(2)$-Lipkin model \cite{6-1}. 
In this argument, a certain $su(2)$-algebra which is independent of the $su(2)$-algebra in the Lipkin model plays a central role. 
We called it as the auxiliary $su(2)$-algebra. 
The orthogonal sets obtained under this algebra give us the minimum weight states of the $su(2)$-Lipkin model. 
We extend this idea to the $su(n)$-Lipkin model. 
Condition that the auxiliary $su(2)$-algebra is independent of the $su(n)$-algebra in the Lipkin model is formulated in the 
commutation relation
\beq\label{1-1}
& &[\ {\rm any\ auxiliary}\ su(2){\rm -generator\ }, \ {\rm any}\ su(n){\rm -generator\ in\ the\ Lipkin\ model\ }]=0\ .
\nonumber\\
& &
\eeq 
For construction of this auxiliary algebra, the raising operator in the $su(2)$-algebra can be expressed in certain form 
with $n$-th degree for the fermion creation operators 
and the Clifford numbers, unfamiliar to nuclear theory. 
The minimum weight states of the $su(n)$-Lipkin model are given 
in terms of the orthogonal sets of the auxiliary $su(2)$-algebra. 
In this paper, the terminology, the ``closed-shell" system was used for the case in which, in the absence of interaction, all 
fermions occupy fully energetically the lowest single-particle level, i.e., $N=2\Omega$. 
However, in order to formulate the ``closed-shell" system rigorously, not only the condition $N=2\Omega$ 
but also other conditions, for example, in the case of the $su(2)$-Lipkin model, $s=\Omega$ ($s$ : the magnitude 
of the $su(2)$-spin for this model) are necessary.

In next section, the $su(n)$-Lipkin model is recapitulated and the condition 
governing the minimum weight states is given. 
In \S 3, the $su(2)$-algebra auxiliary to the $su(n)$-Lipkin model is formulated under the condition 
that any of the $su(2)$-generators commutes with any of the $su(n)$-generators. 
The three generators are expressed as functions of single-particle fermion operators. 
For obtaining the expressions, the Clifford number may be necessary. 
In \S 4, formal aspects of the minimum weight states of the $su(2)$- and the $su(3)$-Lipkin model 
are discussed. 
Section 5 is devoted to presenting the general forms of the minimum weight states concretely in the case of the $su(n)$-Lipkin model. 
Finally, in \S 6, mainly, 
the minimum weight states for the $su(n)$-Lipkin model in the cases $n=2,\ 3,\ 4$ and 5 are given in the form 
slightly different from that presented in {\S 5} 
and it will be useful for the discussion in (II). 

\setcounter{equation}{0}
\section{The $su(n)$-algebra in the Lipkin model}

Many-fermion model discussed in this paper consists of $n$ single-particle levels, the degeneracies of which 
are equal to $2\Omega=2j+1$ ($j$; half-integer). 
The single-particle states are specified by the quantum numbers ($p,jm$). 
Here, $p$ and $m$ are given by $p=0,\ 1,\ 2,\cdots ,\ n-1$ and $m=-j,\ -j+1,\cdots ,\ j-1,\ j$, respectively. 
Hereafter, we omit the quantum number $j$. 
Following the order $p=0<p=1<\cdots <p=n-1$, the levels becomes higher. 
The level $p=0$ is the lowest. 
The fermion operators are denoted by $({\tilde c}_{p,m},\ {\tilde c}_{p,m}^*)$ and, then, 
the total fermion number operator ${\wtilde N}(n)$ for the case $n$ can be expressed as 
\beq\label{2-1}
{\wtilde N}(n)
=\sum_{p=0}^{n-1}\sum_{m=-j}^j{\tilde c}_{p,m}^*{\tilde c}_{p,m}\ . 
\eeq
With the use of the above fermion operators, we can define the following operators for $p,\ q=1,\ 2,\cdots ,\ n-1$:
\bsub\label{2-2}
\beq
& &{\wtilde S}^p(n)=\sum_m{\tilde c}_{p,m}^*{\tilde c}_{0,m}\ , \qquad
{\wtilde S}_p(n)=\sum_m{\tilde c}_{0,m}^*{\tilde c}_{p,m}\ , \quad \left({\wtilde S}_p(n)^*={\wtilde S}^p(n)\right) 
\label{2-2a}\\
& &{\wtilde S}_q^p(n)=\sum_m{\tilde c}_{p,m}^*{\tilde c}_{q,m}-\delta_{pq}\sum_m{\tilde c}_{0,m}^*{\tilde c}_{0,m}\ . \qquad
\left({\wtilde S}_p^q(n)^*={\wtilde S}_q^p(n)\right)
\label{2-2b}
\eeq
\esub
The commutation relations are given in the form 
\bsub\label{2-3}
\beq
& &[\ {\wtilde S}^p(n)\ , \ {\wtilde S}_q(n)\ ]={\wtilde S}_q^p(n)\ , 
\label{2-3a}\\
& &[\ {\wtilde S}_q^p(n)\ , \ {\wtilde S}^r(n)\ ]=\delta_{qr}{\wtilde S}^p(n)+\delta_{pq}{\wtilde S}^r(n)\ , 
\label{2-3b}\\
& &[\ {\wtilde S}_q^p(n)\ , \ {\wtilde S}_r^s(n)\ ]=\delta_{qs}{\wtilde S}_r^p(n)-\delta_{pr}{\wtilde S}_q^s(n)\ . 
\label{2-3c}
\eeq
\esub
In the relation (\ref{2-3}), we can see that the operators (\ref{2-2}) obey the $su(n)$-algebra. 
The simplest Casimir operator, ${\wtilde \Gamma}_{su(n)}$, is given as 
\beq\label{2-4}
& &{\wtilde \Gamma}_{su(n)}=\frac{1}{2}\left[\sum_{p=1}^{n-1}\left({\wtilde S}^p(n){\wtilde S}_p(n)+{\wtilde S}_p(n){\wtilde S}^p(n)\right)
+\sum_{p,q=1}^{n-1}{\wtilde S}_q^p(n){\wtilde S}_p^q(n)-\frac{1}{n}\left(\sum_{p=1}^{n-1}{\wtilde S}_p^p(n)\right)^2\right] . \nonumber\\
& &
\eeq
The operators ${\wtilde \Gamma}_{su(n)}$ and ${\wtilde N}(n)$ satisfy 
\beq\label{2-5}
[\ {\wtilde \Gamma}_{su(n)}\ {\rm and}\ {\wtilde N}(n)\ , \ {\rm any\ of\ the\ operators\ (2.2)}\ ]=0\ .
\eeq
Further, it should be noted that ${\wtilde N}(n)$ can not be expressed in terms of the above $su(n)$-generators.

For the above $su(n)$-algebra, we can select a Hamiltonian 
\beq\label{add2-6}
{\wtilde H}(n)={\wtilde H}_0(n)+{\wtilde H}_1(n)\ . 
\eeq
Here, ${\wtilde H}_0(n)$ is the Hamiltonian of individual levels with energies $\varepsilon_p$, for which we set up 
\beq\label{add2-7}
\sum_{p=0}^{n-1}\varepsilon_p=0\ , \qquad 
\varepsilon_0 \leq \varepsilon_1 \leq \cdots \leq \varepsilon_{n-1}\ .
\eeq
Then, ${\wtilde H}_0(n)$ can be expressed as 
\beq\label{add2-8}
{\wtilde H}_0(n)=\sum_{p=0}^{n-1}\varepsilon_p{\wtilde N}_p(n)=\sum_{p=1}^{n-1}\varepsilon_p{\wtilde S}_p^p(n)\ . 
\eeq
The part ${\wtilde H}_1(n)$ is an interaction term which choose, for illustration only, 
in the form 
\beq\label{add2-9}
{\wtilde H}_1(n)=-G\sum_{p=1}^{n-1}\left[\left({\wtilde S}^p(n)\right)^2+\left({\wtilde S}_p(n)\right)^2\right]\ .
\eeq
Here, $G(>0)$ denotes the coupling constant. 
The above Hamiltonian can be found in Ref.\citen{KM} with the notations different from the present. 
We call the above many-fermion system as the $su(n)$-Lipkin model. 
The Hamiltonian ${\wtilde H}(n)$ obeys
\beq\label{add2-10}
\left[\ {\wtilde \Gamma}_{su(n)}\ {\rm and}\ {\wtilde N}(n)\ , \ {\wtilde H}(n)\ \right]=0\ .
\eeq
The cases $n=2$ and 3 reduce to the Hamiltonians of the $su(2)$- and the $su(3)$-Lipkin model which have been discussed in 
various problems \cite{KM}.

For studies of any many-fermion system, implicitly or explicitly, 
we must prepare orthogonal sets for the system under investigation. 
Standard idea for treating the present model may be, first, to prepare an orthogonal set 
related a chosen minimum weight state. 
The set may be constructed by operating the generators ${\wtilde S}^p(n)$ $(p=1,\ 2,\cdots ,\ n-1)$ and 
${\wtilde S}_q^p(n)\ (p>q=1,\ 2,\cdots ,\ n-2)$ {\it appropriately} on the minimum weight state, 
which we denote $\ket{{\rm min}(n)}$. 
The state $\ket{{\rm min}(n)}$ obeys the conditions 
\beq\label{add2-11}
{\wtilde N}(n)\ket{{\rm min}(n)}=N_{n-1}\ket{{\rm min}(n)}\ , \qquad\qquad\qquad\qquad\ \ 
\eeq
\bsub\label{add2-12}
\beq
& &{\wtilde S}_p(n)\ket{{\rm min}(n)}=0\ , \qquad (p=1,\ 2,\cdots ,\ n-1)
\label{add2-12a}\\
& &{\wtilde S}_p^q(n)\ket{{\rm min}(n)}=0\ , \qquad (p>q=1,\ 2,\cdots ,\ n-2)\qquad\quad\ 
\label{add2-12b}
\eeq
\esub
\beq\label{add2-13}
& &{\wtilde S}_p^p(n)\ket{{\rm min}(n)}=s_p(n)\ket{{\rm min}(n)}\ .  \qquad (p=1,\ 2,\cdots ,\ n-1)
\eeq
Conventionally, for $\ket{{\rm min}(n)}$, a ``closed-shell" system has been investigated:
\beq\label{add2-14}
N_{n-1}=2\Omega\ , \qquad s_p(n)=-2\Omega\ . \qquad (p=1,\ 2,\cdots ,\ n-1)
\eeq
The above teaches us that the level $p=0$ is fully occupied and the levels $p=1,\ 2,\cdots , \ n-1$ are vacant. 
However, even if the treatment is restricted to the ``closed-shell" system, there exist many ``closed-shell" systems 
in the case $n \geq 4$, for example, the levels $p=0$ and 1 are fully occupied and the other vacant: 
\beq\label{add2-15}
N_{n-1}=4\Omega\, \qquad s_{p=1}(n)=0\ , \qquad s_p(n)=-2\Omega\ . \ \  (p=2,\ 3,\cdots ,\ n-1)\ .  \ \ \ 
\eeq
Including such ``closed-shell" systems, it may be interesting to investigate the case with arbitrary fermion number, i.e., 
$0\leq N_{n-1} \leq 2n\Omega$. 
Further, for constructing the orthogonal set built on $\ket{{\rm min}(n)}$, appropriate choice of the operators as functions of 
${\wtilde S}^p(n)\ (p=1,\ 2,\cdots ,\ n-1)$ and ${\wtilde S}_q^p(n)\ (p>q=1,\ 2,\cdots , \ n-2)$ is inevitable. 
The simplest examples are given by 
${\wtilde S}^p(n)\ket{{\rm min}(n)}$ and ${\wtilde S}_q^p(n)\ket{{\rm min}(n)}$. 
However, ${\wtilde S}_q^p(n){\wtilde S}^q(n)\ket{{\rm min}(n)}$ and 
${\wtilde S}^q(n){\wtilde S}_q^p(n)\ket{{\rm min}(n)}$ are not independent of each other, because of the relation 
$[\ {\wtilde S}_q^p(n)\ , \ {\wtilde S}^q(n)\ ]={\wtilde S}^p(n)$. 
We call the operators appropriately chosen as the building blocks. 
The above argument tells us that, as was mentioned in {\S 1}, we have two tasks for formulate the present model: 
(1) One is to determine the minimum weight state and (2) the other is to construct the building blocks. 
Although these two are interrelated with each other, the concrete contents are completely independent of each other. 
Therefore, after discussing the task (1) in (I), we will treat the task (2) in (II).

%
\begin{figure}[t]
\begin{center}
\includegraphics[height=5.5cm]{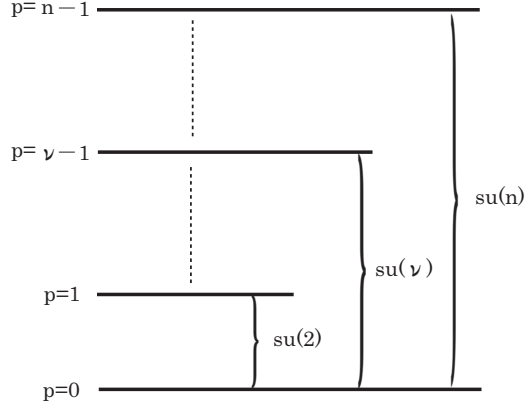}
\caption{The single-particle levels for the $su(n)$-Lipkin models are schematically depicted.
}
\label{fig:fig1}
\end{center}
\end{figure}
%

Main aim of this paper is to present an idea, under which the minimum weight states of the $su(n)$-Lipkin model are systematically constructed. 
In order to make our idea understandable, 
we show the single-particle level scheme in Fig.\ref{fig:fig1}. 
Let $\ket{{\rm min}(\nu)}$ denote a possible candidate of the minimum weight state of the $su(\nu)$-Lipkin model 
for $2\leq \nu \leq n$. 
We set up the following relations for $\ket{{\rm min}(\nu)}$:
\beq\label{2-6}
{\wtilde N}(\nu)\ket{{\rm min}(\nu)}=N_{\nu-1}\ket{{\rm min}(\nu)}\ , 
\qquad\qquad\qquad\qquad\qquad\qquad\qquad
\eeq
\bsub\label{2-7}
\beq
& &{\wtilde S}_p(\nu)\ket{{\rm min}(\nu)}=0\ , \quad (p=1,\ 2,\cdots ,\ \nu-1)
\label{2-7a}\\
& &{\wtilde S}_p^q(\nu)\ket{{\rm min}(\nu)}=0\ , \quad (p>q=1,\ 2, \cdots ,\ \nu-2)
\qquad\qquad\qquad\qquad\qquad
\label{2-7b}
\eeq
\esub
\beq\label{2-8}
& &{\wtilde S}_p^p(\nu)\ket{{\rm min}(\nu)}=(\gamma_{\nu-1}(p)-\gamma_{\nu-1}(0))\ket{{\rm min}(\nu)}\ , \quad (p=1,\ 2,\cdots ,\ \nu-1)
\eeq
Here, $(\gamma_{\nu-1}(p)-\gamma_{\nu-1}(0))$ is given through the relation 
\beq\label{2-9}
\sum_m{\tilde c}_{p,m}^*{\tilde c}_{p,m}\ket{{\rm min}(\nu)}=\gamma_{\nu-1}(p)\ket{{\rm min}(\nu)}\ . 
\quad (p=0,\ 1,\cdots ,\ \nu-1)
\eeq
The total fermion number $N_{\nu-1}$ is expressed as 
\beq\label{2-10}
N_{\nu-1}=\sum_{p=0}^{\nu-1}\gamma_{\nu-1}(p)\ . 
\eeq
It may be necessary to give some comment on the relations (\ref{2-8})$\sim$(\ref{2-10}). 
The definitions of ${\wtilde N}(\nu)$ and ${\wtilde S}_p^p(\nu)$ shown in the relations (\ref{2-1}) and (\ref{2-2b}), respectively, for 
the case $n=\nu$ are rewritten in the form
\beq\label{2-11}
\sum_m{\tilde c}_{p,m}^*{\tilde c}_{p,m}=
\left\{
\begin{array}{llr}
\displaystyle \frac{1}{\nu}\left({\wtilde N}(\nu)-\sum_{q=1}^{\nu-1}{\wtilde S}_q^q(\nu)\right)\ , & (p=0) & (2.21{\rm a})\\
\displaystyle {\wtilde S}_p^p(\nu)+\frac{1}{\nu}\left({\wtilde N}(\nu)-\sum_{q=1}^{\nu-1}{\wtilde S}_q^q(\nu)\right)\ . 
& (p=1,\ 2,\cdots , \ \nu-1) &  (2.21{\rm b})
\end{array}
\right.
\nonumber
\eeq
The relation (\ref{2-11}) tells us the following: 
Since the state $\ket{{\rm min}(\nu)}$ is regarded as the eigenstate of ${\wtilde N}(\nu)$ and ${\wtilde S}_p^p(\nu)$, $\ket{{\rm min}(\nu)}$ 
should be also the eigenstate of $\sum_m{\tilde c}_{0,m}^*{\tilde c}_{0,m}$ and $\sum_m{\tilde c}_{p,m}^*{\tilde c}_{p,m}$. 
Then, the relation (\ref{2-9}) may be permitted to set up and the relations (\ref{2-8}) and (\ref{2-10}) are obtained.

\setcounter{equation}{21}

The relations (\ref{2-6})$\sim$(\ref{2-10}) are set up for the range $2\leq \nu \leq n$. 
However, it may be convenient for later arguments to add the point 
$\nu=1$ to $2\leq \nu \leq n$. 
Judging from Fig.\ref{fig:fig1}, it may be natural to consider that the case 
$\nu=1$ may be restricted only to $p=0$. 
Then, in this case, the relations (\ref{2-7}) and (\ref{2-8}) are meaningless and the relations (\ref{2-6}) and (\ref{2-9}) may be meaningful: 
\bsub\label{2-12}
\beq
{\wtilde N}(1)\ket{{\rm min}(1)}=N_0\ket{{\rm min}(1)}=\gamma_0(0)\ket{{\rm min}(1)}\ . 
\label{2-12a}
\eeq
Here, ${\wtilde N}(1)$ is given by the relation (\ref{2-1}) for $n=1$ in the form
\beq
{\wtilde N}(1)=\sum_m{\tilde c}_{0,m}^*{\tilde c}_{0,m}\ . 
\label{2-12b}
\eeq
\esub

Let $\ket{{\rm min}(\nu)}$ be obtained. 
Then, we can show that $\ket{{\rm min}(\nu)}$ for $\nu=2,\ 3,\cdots ,\ n$ satisfies the relation 
\beq\label{2-13}
{\wtilde N}(n)\ket{{\rm min}(\nu)}=N_{\nu-1}\ket{{\rm min}(\nu)}\ . 
\qquad\qquad\qquad\qquad\qquad
\eeq
\bsub\label{2-14}
\beq
& &{\wtilde S}_p(n)\ket{{\rm min}(\nu)}=0\ , \quad (p=1,\ 2, \cdots ,\ n-1)
\label{2-14a}\\
& &{\wtilde S}_p^q(n)\ket{{\rm min}(\nu)}=0\ , \quad
(p>q=1,\ 2,\cdots , \ n-2)
\qquad\qquad\qquad\qquad\qquad\qquad\qquad\qquad
\label{2-14b}
\eeq
\esub
\beq\label{2-15}
{\wtilde S}_p^p(n)\ket{{\rm min}(\nu)}=
\left\{
\begin{array}{llr}
\displaystyle (\gamma_{\nu-1}(p)-\gamma_{\nu-1}(0))\ket{{\rm min}(\nu)}\ ,  & (p=1,\ 2,\cdots ,\ \nu-1)) & (2.25{\rm a})\\
\displaystyle -\gamma_{\nu-1}(0)\ket{{\rm min}(\nu)}\ . & (p=\nu,\ \nu+1,\cdots , \ n-1) &  (2.25{\rm b})
\end{array}
\right.
\nonumber
\eeq
\setcounter{equation}{25}
The reason is very simple. 
Since any fermion does not occupy the single-particle levels $p=\nu,\ \nu+1,\cdots ,\ n-1$, we have 
\beq\label{2-16}
{\tilde c}_{p,m}\ket{{\rm min}(\nu)}=0\ . \quad (p=\nu,\ \nu+1,\cdots ,\ n-1)
\eeq

The relations (\ref{2-13})$\sim$(\ref{2-15}) teach us that $\ket{{\rm min}(\nu)}$ as the solution 
of Eqs.(\ref{2-6})$\sim$(\ref{2-8}) is also the minimum weight state of the $su(n)$-Lipkin model. 
In next section, we will discuss the $su(2)$-algebra $({\wtilde \Lambda}_{\pm,0}(n)$), 
which plays a central role for obtaining the state $\ket{{\rm min}(\nu)}$.

\setcounter{equation}{0}
\section{The $su(2)$-algebra auxiliary to the $su(n)$-Lipkin model}

As was mentioned in {\bf 1}, an idea preliminary to the present one has been already shown in our recent paper for the case 
of the $su(2)$-Lipkin model \cite{6-1}. 
The basic idea is to introduce the $su(2)$-algebra $({\wtilde \Lambda}_{\pm,0}(2))$, which is characterized 
by the commutation relation 
\beq\label{3-1}
& &\left[\ {\rm any\ of\ }{\wtilde \Lambda}_{\pm,0}(2)\ , \ {\rm any\ of\ the}\ su(2){\rm -generators}\ 
\left({\wtilde S}^1(2),\ {\wtilde S}_1(2),\ {\wtilde S}_1^1(2)\right)\ \right]=0\ . \nonumber\\
& &
\eeq
The explicit forms are as follows:
\bsub\label{3-2}
\beq
& &{\wtilde \Lambda}_+(2)=\sum_m{\tilde c}_{1,m}^*{\tilde c}_{0,m}^*\ , \qquad
{\wtilde \Lambda}_{-}(2)=\sum_m{\tilde c}_{0,m}{\tilde c}_{1,m}\ , 
\label{3-2a}\\
& &{\wtilde \Lambda}_0(2)=\frac{1}{2}\sum_m({\tilde c}_{1,m}^*{\tilde c}_{1,m}+{\tilde c}_{0,m}^*{\tilde c}_{0,m})-\Omega\ \left(=
\frac{1}{2}{\wtilde N}(2)-\Omega\right)\ . 
\label{3-2b}
\eeq
\esub
It is easily verified that the expression (\ref{3-2}) satisfies the condition (\ref{3-1}) and obeys 
the $su(2)$-algebra: 
\beq\label{3-3}
[\ {\wtilde \Lambda}_+(2)\ , \ {\wtilde \Lambda}_-(2)\ ]=2{\wtilde \Lambda}_0(2)\ , \qquad
[\ {\wtilde \Lambda}_0(2)\ , \ {\wtilde \Lambda}_{\pm}(2)\ ]=\pm {\wtilde \Lambda}_{\pm}(2)\ . 
\eeq
In our idea, $({\wtilde \Lambda}_{\pm,0}(2))$ plays a central role in deriving the minimum weight state with arbitrary fermion number 
in the $su(2)$-Lipkin model. 
Conventionally, only the case of the fermion number $2\Omega$ has been treated, i.e., 
the ``closed-shell" system. 
In {\S 4}, for illustration of our idea, we will discuss how $({\wtilde \Lambda}_{\pm,0}(2))$ is used in our present problem including the case of the 
$su(3)$-Lipkin model. 
In the form similar to the relation (\ref{3-2}), we can give the $su(2)$-algebra $({\wtilde \Lambda}_{\pm,0}(n))$ which 
is independent of the $su(n)$-Lipkin model: 
\beq\label{3-4}
\left[\ {\rm any\ of\ }{\wtilde \Lambda}_{\pm,0}(n)\ , \ {\rm any\ of}\ \left({\wtilde S}^p(n),\ {\wtilde S}_p(n),\ {\wtilde S}_q^p(n)\right)\ \right]=0\ . 
\eeq

For constructing $({\wtilde \Lambda}_{\pm,0}(n))$, first, we must have a preliminary argument. 
We know that system composed of one kind of fermion is regarded as single $su(2)$-spin system with 
the magnitude $1/2$. 
Through the following commutation relation, we can understand this point:
\beq\label{3-5}
[\ {\tilde c}^*\ , \ {\tilde c}\ ]=2\left({\tilde c}^*{\tilde c}-\frac{1}{2}\right)\ , \qquad
\left[\ {\tilde c}^*{\tilde c}-\frac{1}{2}\ , \ {\tilde c}^*\ \right]={\tilde c}^*\ , \qquad
({\tilde c}^*)^2=0
\eeq
Here, $({\tilde c}^*, {\tilde c})$ denotes fermion operator obeying the anti-commutation relation
\beq\label{3-6}
\{\ {\tilde c}^*\ , \ {\tilde c}\ \}=1\ , \qquad
\{\ {\tilde c}^*\ , \ {\tilde c}^*\ \}=0\ . 
\eeq
The anti-commutation relation (\ref{3-6}) leads us to the relation (\ref{3-5}). 
The fermion operators ${\tilde c}^*$ and ${\tilde c}$ play a role of the raising and the lowering 
operator, respectively. 
However, the form (\ref{3-5}) cannot be straightforwardly translated into the case of many-fermion system, for example, the system 
specified by $p=0$ in this paper:
\beq\label{3-7}
\{\ {\tilde c}_{0,m}^*\ , \ {\tilde c}_{0,\mu}\ \}=\delta_{m\mu}\ , \qquad
\{\ {\tilde c}_{0,m}^*\ , \ {\tilde c}_{0,\mu}^*\ \}=0\ , \quad {\rm i.e.}\quad ({\tilde c}_{0,m})^2=0\ . 
\eeq
The first of the relation (\ref{3-7}) is rewritten to 
\beq\label{3-8}
[\ {\tilde c}_{0,m}^*\ , \ {\tilde c}_{0,\mu}\ ]=2\left({\tilde c}_{0,m}^*{\tilde c}_{0,\mu}-\frac{1}{2}\delta_{m\mu}\right)\ . 
\eeq
The form (\ref{3-8}) suggests us that it may be impossible to regard ${\tilde c}_{0,m}^*$ 
as the raising operator of many $su(2)$-spin system as it stands.

Let us discuss a possible idea for the above problem. 
Under this idea, the present many-fermion system can be regarded as that composed of independent $2\Omega$ $su(2)$-spins. 
Each is specified by $m$ and its magnitude is equal to $1/2$. 
This idea is realized through introducing the Clifford numbers $e_m\ (m=-j,\ -j+1,\cdots ,\ j-1,\ j)$ which obey the condition 
\beq\label{3-9}
& &e_me_\mu+e_\mu e_m=0\quad {\rm for}\quad m\neq \mu\ , \quad (e_m)^2=1\ , \qquad 
{\rm i.e.,}\quad 
\{\ e_m\ , \ e_{\mu}\ \}=2\delta_{m\mu}\ , 
\nonumber\\
& &
[\ e_m\ , \ {\tilde c}_{0,\mu}^*\ {\rm and}\ {\tilde c}_{0,\mu}\ ]=0\ . 
\eeq 
Of course, $e_m$ commutes with the fermion operators. 
With the use of $e_m$, we define the following operators: 
\beq\label{3-10}
{\tilde d}_{0,m}^*=e_m{\tilde c}_{0,m}^*\ , \qquad
{\tilde d}_{0,m}=e_m{\tilde c}_{0,m}\ . 
\eeq
With the aid of the anti-commutation relation (\ref{3-7}) and the property of the Clifford number (\ref{3-9}), we can derive 
the following relation\!\!
\footnote{
Equation (\ref{3-11b}) can be derived through the following process: \\
\beq
[\ {\tilde d}_{0,m}^*\ , \ {\tilde d}_{0,\mu}\ ] 
&=&e_m{\tilde c}_{0,m}^*\cdot e_{\mu}{\tilde c}_{0,\mu}-e_{\mu}{\tilde c}_{0,\mu}\cdot e_m{\tilde c}_{0,m}^*\nonumber\\
&=&e_me_{\mu}\cdot {\tilde c}_{0,m}^*{\tilde c}_{0,\mu}-e_{\mu}e_m\cdot (\delta_{m\mu}-{\tilde c}_{0,m}^*{\tilde c}_{0,\mu})\nonumber\\
&=&\{\ e_m\ , \ e_{\mu}\ \}\cdot{\tilde c}_{0,m}^*{\tilde c}_{0,\mu}-e_m^2\cdot\delta_{m\mu}
=\delta_{m\mu}\cdot 2\left({\tilde c}_{0,m}^*{\tilde c}_{0,\mu}-\frac{1}{2}\right) \nonumber\\
&=&\delta_{m\mu}\cdot 2\left({\tilde d}_{0,m}^*{\tilde d}_{0,m}-\frac{1}{2}\right)\ .
\nonumber
\eeq
}
for $({\tilde d}_{0,m}^*,{\tilde d}_{0,m})$:
\bsub\label{3-11}
\beq
& &[\ {\tilde d}_{0,m}^*\ , \ {\tilde d}_{0,\mu}^*\ ]=0\ , \qquad
({\tilde d}_{0,m}^*)^2=0\ , 
\label{3-11a}\\
& &[\ {\tilde d}_{0,m}^*\ , \ {\tilde d}_{0,\mu}\ ]=\delta_{m\mu}\cdot 2\left({\tilde d}_{0,m}^*{\tilde d}_{0,m}-\frac{1}{2}\right)
\label{3-11b}\\
& &\left[\ {\tilde d}_{0,m}^*{\tilde d}_{0,m}-\frac{1}{2}\ , \ {\tilde d}_{0,\mu}^*\ \right]=\delta_{m\mu}\cdot{\tilde d}_{0,\mu}^*\ . 
\label{3-11c}
\eeq
\esub
In contrast to the form (\ref{3-8}), we can see that the symbol $\delta_{m\mu}$ is attached 
to both of the two terms on the right-hand side of the relation (\ref{3-11b}). 
Therefore, the relation (\ref{3-11}) suggests us that the present many-fermion system 
consists of $2\Omega$ $su(2)$-spins which are independent of one other and the generators of the $m$-th spin are given by 
$({\tilde d}_{0,m}^*,{\tilde d}_{0,m},{\tilde d}_{0,m}^*{\tilde d}_{0,m}-1/2)$. 
The total spin of the present system, $({\wtilde \Lambda}_{\pm,0}(1))$, can be expressed in the form 
\bsub\label{3-12}
\beq
& &{\wtilde \Lambda}_+(1)=\sum_m{\tilde d}_{0,m}^*\ \left(=\sum_me_m{\tilde c}_{0,m}^*\right)\ , \qquad
{\wtilde \Lambda}_-(1)=\sum_m{\tilde d}_{0,m}\ \left(=\sum_me_m{\tilde c}_{0,m}\right)\ , 
\label{3-12a}\\
& &{\wtilde \Lambda}_0(1)=\sum_m\left({\tilde d}_{0,m}^*{\tilde d}_{0,m}-\frac{1}{2}\right)\ 
\left(=\sum_m{\tilde c}_{0,m}^*{\tilde c}_{0,m}-\Omega={\wtilde N}(1)-\Omega\right)\ .\quad (e_m^2=1)
\qquad\qquad
\eeq
\esub
Of course, they obey the $su(2)$-algebra:
\beq\label{3-13}
[\ {\wtilde \Lambda}_+(1)\ , \ {\wtilde \Lambda}_-(1)\ ]=2{\wtilde \Lambda}_0(1)\ , \qquad
[\ {\wtilde \Lambda}_0(1)\ , \ {\wtilde \Lambda}_{\pm}(1)\ ]=\pm{\wtilde \Lambda}_{\pm}(1)\ . 
\eeq
We can treat the eigenvalue problem of $({\wtilde \Lambda}_{\pm,0}(1))$, which will be discussed in {\bf 4} 
in reference to the $su(2)$- and the $su(3)$-Lipkin model. 
The $su(2)$-algebra $({\tilde \Lambda}_{\pm,0}(2))$ given in the relation (\ref{3-2}) is expressed as 
\bsub\label{3-14}
\beq
& &{\wtilde \Lambda}_+(2)=\sum_m{\tilde d}_{1,m}^*{\tilde d}_{0,m}^*\ , \qquad
{\wtilde \Lambda}_-(2)=\sum_m{\tilde d}_{0,m}{\tilde d}_{1,m}\ , 
\label{3-14a}\\
& &{\wtilde \Lambda}_0(2)=\frac{1}{2}\sum_m({\tilde d}_{1,m}^*{\tilde d}_{1,m}+{\tilde d}_{0,m}^*{\tilde d}_{0,m})-\Omega\ . 
\label{3-14b}
\eeq
\esub
Here, we used $(e_m)^2=1$ and $({\tilde d}_{0,m}^*,{\tilde d}_{0,m})$ and $({\tilde d}_{1,m}^*,{\tilde d}_{1,m})$ are given through 
\beq\label{3-15}
{\tilde d}_{p,m}^*=e_m{\tilde c}_{p,m}^*\ , \qquad
{\tilde d}_{p,m}=e_m{\tilde c}_{p,m}\ . \quad (p=0,\ 1,\cdots ,\ n-1)
\eeq
\setcounter{equation}{14}
Properties of the above operators are summarized as follows
\!\!
\footnote{
The second of the relation (\ref{3-15a}) can be derived through the following process: 
\beq
\{\ {\tilde d}_{p,m}\ , \ {\tilde d}_{q,m}^*\ \}
&=&e_m{\tilde c}_{p,m}\cdot e_m{\tilde c}_{q,m}^*+e_m{\tilde c}_{q,m}^*\cdot e_m{\tilde c}_{p,m}\nonumber\\
&=&(e_m)^2\{\ {\tilde c}_{p,m}\ , \ {\tilde c}_{q,m}^*\ \}=\delta_{pq}\ . 
\nonumber
\eeq
}
: 
\bsub\label{3-15add}
\beq
& &\{\ {\tilde d}_{p,m}^*\ , \ {\tilde d}_{q,\mu}^*\ \}=0\ , \qquad
\{\ {\tilde d}_{p,m}\ , \ {\tilde d}_{q,\mu}^*\ \}=\delta_{pq}\quad {\rm for}\quad m=\mu\ , 
\label{3-15a}\\
& &[\ {\tilde d}_{p,m}^*\ , \ {\tilde d}_{q,\mu}^*\ ]=0\ , \qquad
[\ {\tilde d}_{p,m}\ , \ {\tilde d}_{q,\mu}^*\ ]=0 \quad {\rm for}\quad m\neq\mu\ . 
\label{3-15b}
\eeq
\esub

We are now possible to give explicit forms ${\wtilde \Lambda}_{\pm,0}(n)$. 
First, we define the following operators:
\beq\label{3-16}
{\tilde d}_{m}^*(n)={\tilde d}_{n-1,m}^*{\tilde d}_{n-2,m}^*\cdots {\tilde d}_{1,m}^*{\tilde d}_{0,m}^*\ , 
\eeq
i.e.,
\beq\label{3-17}
{\tilde d}_m^*(n)=
\left\{
\begin{array}{llr}
{\tilde c}_{n-1,m}^*{\tilde c}_{n-2,m}^*\cdots {\tilde c}_{1,m}^*{\tilde c}_{0,m}^* & {\rm for}\ n:{\rm even}\ ((e_m)^n=1)\ ,& (3.17{\rm a}) \\
e_m{\tilde c}_{n-1,m}^*{\tilde c}_{n-2,m}^*\cdots {\tilde c}_{1,m}^*{\tilde c}_{0,m}^* & {\rm for}\ n:{\rm odd}\ ((e_m)^n=e_m)\ ,& 
\qquad\quad 
(3.17{\rm b})
\end{array}
\right.
\nonumber
\eeq
\setcounter{equation}{17}
Clearly, ${\tilde d}_m^*(1)={\tilde d}_{0,m}^*$ and ${\tilde d}_m^*(2)={\tilde d}_{1,m}^*{\tilde d}_{0,m}^*$, which were used in the 
expressions (\ref{3-12}) and (\ref{3-14}), respectively. 
The operators $({\tilde d}_m^*(n),{\tilde d}_m(n))$ satisfy the relation
\bsub\label{3-18}
\beq
& &{\tilde d}_m^*(n)={\tilde d}_m^*(n)\cdot{\tilde d}_m(n)\cdot{\tilde d}_m^*(n)\ , 
\label{3-18a}\\
& &({\tilde d}_m^*(n))^2=0\ . 
\label{3-18b}
\eeq
\esub
The above two relations are compatible with each other. 
Further, we have 
\bsub\label{3-19}
\beq
& &[\ {\tilde d}_m^*(n)\ , \ {\tilde d}_\mu^*(n)\ ]=0\quad {\rm for\ any\ combination\ of}\ (m,\mu)\ , 
\label{3-19a}\\
& &[\ {\tilde d}_m^*(n)\ , \ {\tilde d}_{\mu}(n)\ ]=0\quad {\rm for}\ m\neq \mu\ .
\label{3-19b}
\eeq
\esub
Judging from the expressions (\ref{3-12}) and (\ref{3-14}), it may be natural to set up the 
following form for $({\wtilde \Lambda}_{\pm,0}(n))$:
\bsub\label{3-20}
\beq
& &{\wtilde \Lambda}_+(n)=\sum_m{\tilde d}_m^*(n)\ , \qquad
{\wtilde \Lambda}_-(n)=\sum_m{\tilde d}_m(n)\ , 
\label{3-20a}\\
& &{\wtilde \Lambda}_0(n)=\frac{1}{2}\sum_m[\ {\tilde d}_m^*(n)\ , \ {\tilde d}_m(n)\ ]\ . 
\label{3-20b}
\eeq
\esub
It should be noted that the $su(2)$-algebra $({\wtilde \Lambda}_{\pm,0}(n))$ is extended from the fermion-pair for $p=0$ and 1 
$({\wtilde \Lambda}_{\pm,0}(2))$. 
With the use of the relations (\ref{3-18}) and (\ref{3-19}), we can show that ${\wtilde \Lambda}_{\pm,0}(n)$ obey the $su(2)$-algebra: 
\beq\label{3-21}
[\ {\wtilde \Lambda}_+(n)\ , \ {\wtilde \Lambda}_-(n)\ ]=2{\wtilde \Lambda}_0(n)\ , \qquad
[ \ {\wtilde \Lambda}_0(n)\ , \ {\wtilde \Lambda}_{\pm}(n)\ ]=\pm{\wtilde \Lambda}_{\pm}(n)\ . 
\eeq

Next, we will give the proof of the commutation relation (\ref{3-4}). 
For this aim, we express the $su(n)$-generators (\ref{2-2}) in the unified form 
\beq\label{3-22}
{\wtilde S}_{\sigma}^{\rho}(n)=
\sum_m({\tilde c}_{\rho,m}^*{\tilde c}_{\sigma, m}-\delta_{\rho\sigma}{\tilde c}_{0,m}^*{\tilde c}_{0,m})\ . \qquad
(\rho,\ \sigma=0,\ 1,\cdots,\ n-2,\ n-1)\qquad
\eeq
Of course, ${\wtilde S}_0^0(n)=0$. 
On the other hand, picking up ${\tilde c}_{\rho,m}^*{\tilde c}_{\sigma,m}$, ${\wtilde \Lambda}_+(n)$ shown 
in the relation (\ref{3-20}) with (\ref{3-17}) can be factorized as follows: 
\beq\label{3-23}
{\wtilde \Lambda}_+(n)=\sum_m{\wtilde \Lambda}_m^{(+)}(n;\rho\sigma)\cdot{\tilde c}_{\rho,m}^*{\tilde c}_{\sigma,m}^*\ . 
\eeq
It should be noted that ${\wtilde \Lambda}_m^{(+)}(n;\rho\sigma)$ does not contain ${\tilde c}_{\rho,m}^*{\tilde c}_{\sigma,m}^*$. 
Then, for $\rho\neq \sigma$, we have 
\bsub\label{3-24}
\beq
& &[\ {\wtilde \Lambda}_+(n)\ , \ {\wtilde S}_\sigma^\rho (n)\ ]=\sum_m{\wtilde \Lambda}_m^{(+)}(n;\rho\sigma)
[\ {\tilde c}_{\rho,m}^*{\tilde c}_{\sigma,m}^*\ , \ {\tilde c}_{\rho,m}^*{\tilde c}_{\sigma,m}\ ]=0\ , 
\label{3-24a}\\
& &[\ {\wtilde \Lambda}_-(n)\ , \ {\wtilde S}_\sigma^\rho(n)\ ]=-[\ {\wtilde \Lambda}_+(n)\ , \ {\wtilde S}_\rho^\sigma (n)\ ]^*=0\ . 
\label{3-24b}
\eeq
\esub
For the case $\rho=\sigma$, we have 
\bsub\label{3-25}
\beq
& &[\ {\wtilde \Lambda}_+(n)\ , \ {\wtilde S}_\rho^\rho (n)\ ]=\sum_m{\wtilde \Lambda}_m^{(+)}(n;\rho\sigma=0)
[\ {\tilde c}_{\rho,m}^*{\tilde c}_{0,m}^*\ , \ {\tilde c}_{\rho,m}^*{\tilde c}_{\rho,m}-{\tilde c}_{0,m}^*{\tilde c}_{0,m}\ ]=0\ , \qquad\quad
\label{3-25a}\\
& &[\ {\wtilde \Lambda}_-(n)\ , \ {\wtilde S}_\rho^\rho(n)\ ]=-[\ {\wtilde \Lambda}_+(n)\ , \ {\wtilde S}_\rho^\rho (n)\ ]^*=0\ . 
\label{3-25b}
\eeq
\esub
The relation ${\wtilde \Lambda}_0(n)=[\ {\wtilde \Lambda}_+(n)\ , \ {\wtilde \Lambda}_-(n)\ ]/2$ gives us 
\beq\label{3-26}
[\ {\wtilde \Lambda}_0(n)\ , \ {\wtilde S}_{\sigma}^{\rho}(n)\ ]=0\ . 
\eeq
In this way, we could show that the expression (\ref{3-20}) satisfies the relation (\ref{3-4}).

In next section, the expressions of ${\wtilde \Lambda}_{\pm,0}(2)$ shown in the relation (\ref{3-14}) and 
${\wtilde \Lambda}_{\pm,0}(3)$ shown in the following play a central role: 
\bsub\label{3-27}
\beq
{\wtilde \Lambda}_+(3)&=&
\sum_m e_m{\tilde c}_{2,m}^*{\tilde c}_{1,m}^*{\tilde c}_{0,m}^*\ , \qquad
{\wtilde \Lambda}_-(3)=
\sum_m e_m{\tilde c}_{0,m}{\tilde c}_{1,m}{\tilde c}_{2,m}\ , 
\label{3-27a}\\
{\wtilde \Lambda}_0(3)&=&
\frac{1}{2}\sum_m({\tilde c}_{2,m}^*{\tilde c}_{2,m}+{\tilde c}_{1,m}^*{\tilde c}_{1,m}+{\tilde c}_{0,m}^*{\tilde c}_{0,m}) \nonumber\\
&-&\frac{1}{2}\sum_m({\tilde c}_{2,m}^*{\tilde c}_{2,m}\cdot{\tilde c}_{1,m}^*{\tilde c}_{1,m}
+{\tilde c}_{1,m}^*{\tilde c}_{1,m}\cdot{\tilde c}_{0,m}^*{\tilde c}_{0,m}
+{\tilde c}_{0,m}^*{\tilde c}_{0,m}\cdot{\tilde c}_{2,m}^*{\tilde c}_{2,m})
\nonumber\\
&+&\sum_m{\tilde c}_{2,m}^*{\tilde c}_{2,m}\cdot{\tilde c}_{1,m}^*{\tilde c}_{1,m}\cdot{\tilde c}_{0,m}^*{\tilde c}_{0,m}-\Omega\ . 
\label{3-27b}
\eeq
\esub

\setcounter{equation}{0}
\section{The minimum weight states of the $su(2)$- and the $su(3)$-Lipkin model}

In order to illustrate our idea, let us start with the $su(2)$-Lipkin model. 
We denote one of the states in which only the single-particle level $p=0$ is occupied by $N_0$ fermions as $\ket{N_0}$:
\beq\label{4-1}
{\wtilde N}(1)\ket{N_0}=N_0\ket{N_0}\ , \quad {\rm i.e.,}\quad
{\wtilde N}(2)\ket{N_0}=N_0\ket{N_0}\ . 
\eeq
Here, we omitted any quantum number which does not connect with the algebras under consideration. 
It is easily verified that $\ket{N_0}$ is a possible candidate of the minimum weight states of the $su(2)$-Lipkin model: 
\beq\label{4-2}
{\wtilde S}_1(2)\ket{N_0}=0\ , \qquad {\wtilde S}_1^1(2)\ket{N_0}=-N_0\ket{N_0}\ . \quad (N_0\geq 0)
\eeq
Comparison of the relations (\ref{4-1}) and (\ref{4-2}) with (\ref{2-14}), (2.15a) and (\ref{2-16}) gives us, 
for the case $(n=2,\ \nu=1,\ p=1)$:
\beq\label{4-3}
\ket{{\rm min}(1)}=\ket{N_0}\ , \qquad \gamma_0(0)=N_0\ . 
\eeq
An example of $\ket{N_0}$ is presented in Appendix.

The state $\ket{N_0}$ is also the minimum weight state of the $su(2)$-algebra $({\wtilde \Lambda}_{\pm,0}(2))$:
\bsub\label{4-13}
\beq
& &{\wtilde \Lambda}_-(2)\ket{N_0}=0\ , 
\label{4-13a}\\
& &
{\wtilde \Lambda}_0(2)\ket{N_0}=-\lambda(2)\ket{N_0}\ , \qquad
\lambda(2)=\Omega-\frac{N_0}{2}\ . 
\label{4-13b} 
\eeq
\esub
Therefore, by operating ${\wtilde \Lambda}_+(2)$ successively on $\ket{N_0}$, we are able to obtain the states orthogonal to $\ket{N_0}$ 
in the form 
\beq\label{4-14}
\ket{N_1,N_0}=\left({\wtilde \Lambda}_+(2)\right)^{\frac{N_1-N_0}{2}}\ket{N_0}\ . \quad
(0\leq N_0 \leq N_1)
\eeq
The state $\ket{N_1,N_0}$ satisfies 
\beq
& &{\wtilde N}(2)\ket{N_1,N_0}=N_1\ket{N_1,N_0}\ , 
\label{4-15}\\
& &{\wtilde S}_1(2)\ket{N_1,N_0}=0\ , \qquad
{\wtilde S}_1^1(2)\ket{N_1,N_0}=-N_0\ket{N_1,N_0}\ , \ \ \ \ \ \qquad
\label{4-16}
\eeq
\bsub\label{4-17}
\beq
& &{\wtilde \Lambda}_-(2)\ket{N_1,N_0}\neq 0\ , 
\label{4-17a}\\
& &{\wtilde \Lambda}_0(2)\ket{N_1,N_0}=\lambda_0(2)\ket{N_1,N_0}\ , \qquad
\lambda_0(2)=\frac{N_1-N_0}{2}-\lambda(2)\ . 
\label{4-17b}
\eeq
\esub
The state $\ket{N_1,N_0}$ is also the minimum weight state of the $su(2)$-Lipkin model with the same property as that shown in the relation 
(\ref{4-2}). 
But, it is not the minimum weight state of the $su(2)$-algebra $({\wtilde \Lambda}_{\pm,0}(2))$. 
For $\ket{{\rm min}(2)}=\ket{N_1,N_0}$, we obtain the following: 
\bsub\label{4-18}
\beq
& &\gamma_1(0)=\frac{N_1-N_0}{2}+N_0\ , \qquad
\gamma_1(1)=\frac{N_1-N_0}{2}\ . 
\label{4-18a}
\eeq
Inversely, we have 
\beq
N_0=\gamma_1(0)-\gamma_1(1)\ , \qquad
N_1=\gamma_1(0)+\gamma_1(1)\ . 
\label{4-18b}
\eeq
\esub
The above relations lead us to the inequalities 
\beq\label{4-19}
0\leq N_0 \leq N_1\ , \qquad
0\leq \gamma_1(1) \leq \gamma_1(0)\ . 
\eeq
Since ${\wtilde \Lambda}_{\pm,0}(2)$ obey the $su(2)$-algebra, the relations (\ref{4-13b}) and (\ref{4-17b}) give us the following inequalities:
\bsub\label{4-20}
\beq
& &0\leq \Omega-\frac{N_0}{2}\ , \quad{\rm i.e.,}\quad 0\leq N_0 \leq 2\Omega\ , 
\label{4-20a}\\
& &-\left(\Omega-\frac{N_0}{2}\right) \leq -\left(\Omega-\frac{N_1}{2}\right) \leq \Omega-\frac{N_0}{2}\ , \quad
{\rm i.e.,}\quad 
0\leq N_0 \leq N_1 \leq 4\Omega-N_0\ . \qquad
\label{4-20b}
\eeq
\esub
Fermion numbers in the single-particle levels $p=0$ and $p=1$ are given in the relation (\ref{4-19}) and, 
then, we have 
\beq\label{4-21}
0\leq \gamma_1(1) \leq \gamma_1(0)\leq 2\Omega\ . 
\eeq
Of course, if $N_1=N_0$, $\gamma_0(1)=N_0$ and $\gamma_1(1)=0$. 
The above is an outline of the $su(2)$-Lipkin model based on the present idea and, 
needless to say, it is consistent to the result shown in our recent work. 
We were able to obtain the minimum weight states of the Lipkin model with any fermion numbers 
governed by the condition (\ref{4-21}).

Next, we consider the minimum weight states of the $su(3)$-Lipkin model. 
First, we pay an attention to the state $\ket{N_1,N_0}$ shown in the relation (\ref{4-14}), which satisfies 
\beq
& &{\wtilde N}(3)\ket{N_1,N_0}=N_1\ket{N_1,N_0}\ , 
\label{4-22}\\
& &{\wtilde S}_1(3)\ket{N_1,N_0}={\wtilde S}_2(3)\ket{N_1,N_0}={\wtilde S}_2^1(3)\ket{N_1,N_0}=0\ , 
\label{4-23}\\
& &{\wtilde S}_1^1(3)\ket{N_1,N_0}=-N_0\ket{N_1,N_0}\ , \qquad
{\wtilde S}_2^2(3)\ket{N_1,N_0}=-\frac{1}{2}(N_1+N_0)\ket{N_1,N_0}\ . \ 
\label{4-24}
\eeq 
For the relation (\ref{4-23}), we should note that ${\wtilde S}_1(3)={\wtilde S}_1(2)$ and, further, 
$\ket{N_1,N_0}$ does not contain any fermion in the level $p=2$ and 
${\wtilde S}_2(3)$ and ${\wtilde S}_2^1(3)$ contain the annihilation operator in $p=2$. 
Although $\ket{N_1,N_0}$ is not minimum weight state of $({\wtilde \Lambda}_{\pm,0}(2))$, 
it is the minimum weight state of $({\wtilde \Lambda}_{\pm,0}(3))$: 
\bsub\label{4-25}
\beq
& &{\wtilde \Lambda}_-(3)\ket{N_1,N_0}=0\ , 
\label{4-25a}\\
& &{\wtilde \Lambda}_0(3)\ket{N_1, N_0}=-\lambda(3)\ket{N_1,N_0}\ , 
\qquad
\lambda(3)=\Omega-\frac{1}{2}\left(\frac{N_1-N_0}{2}+N_0\right)\ .  
\label{4-25b}
\eeq
\esub
If $N_1=N_0$, $\ket{N_0}\ (=\ket{N_1=N_0,N_0})$ is also the minimum weight state of the $su(3)$-Lipkin model. 
It may be clear from the relations (\ref{4-22})$\sim$(\ref{4-25}). 
Then, we introduce the state $\ket{N_2,N_1,N_0}$ in the form 
\beq\label{4-26}
\ket{N_2,N_1,N_0}=\left({\wtilde \Lambda}_+(3)\right)^{\frac{N_2-N_1}{3}}\ket{N_1,N_0}\ . 
\eeq
The state $\ket{N_2,N_1,N_0}$ satisfies 
\beq
& &{\wtilde N}(3)\ket{N_2,N_1,N_0}=N_2\ket{N_2,N_1,N_0}\ , 
\label{4-27}\\
& &{\wtilde S}_1(3)\ket{N_2,N_1,N_0}={\wtilde S}_2(3)\ket{N_2,N_1,N_0}
={\wtilde S}_2^1(3)\ket{N_2,N_1,N_0}=0\ , 
\label{4-28}\\
& &
{\wtilde S}_1^1(3)\ket{N_2,N_1,N_0}=-N_0\ket{N_2,N_1,N_0}\ , \nonumber\\ 
& &{\wtilde S}_2^2(3)\ket{N_2,N_1,N_0}=-\frac{1}{2}(N_1+N_0)\ket{N_2,N_1,N_0} , 
\label{4-29}
\eeq
\bsub\label{4-30}\beq
& &{\wtilde \Lambda}_-(3)\ket{N_2,N_1,N_0}\neq 0\ , 
\label{4-30a}\\
& &{\wtilde \Lambda}_0(3)\ket{N_2,N_1,N_0}=\lambda_0(3)\ket{N_2,N_1,N_0}\ ,\qquad 
\lambda_0(3)=\frac{N_2-N_1}{3}-\lambda(3)\ . \qquad\ \ \ 
\label{4-30b}
\eeq
\esub
The state $\ket{N_2,N_1,N_0}$ is also the minimum weight state of the $su(3)$-Lipkin model with the same property as that shown 
in the relation (\ref{4-23}) and (\ref{4-24}). 
But, it is not the minimum weight state of the $su(2)$-algebra $({\wtilde \Lambda}_{\pm,0}(3))$. 
For $\ket{{\rm min}(3)}=\ket{N_2,N_1,N_0}$, we obtain the following:
\bsub\label{4-31}
\beq
& &\gamma_2(0)=\frac{N_2-N_1}{3}+\frac{N_1-N_0}{2}+N_0\ , \nonumber\\
& &\gamma_2(1)=\frac{N_2-N_1}{3}+\frac{N_1-N_0}{2}\ , \nonumber\\
& &\gamma_2(2)=\frac{N_2-N_1}{3}\ . 
\label{4-31a}
\eeq
Inversely, we have 
\beq
& &N_0=\gamma_2(0)-\gamma_2(1)\ , \quad
N_1=\gamma_2(0)+\gamma_2(1)-2\gamma_2(2)\ , \quad
N_2=\gamma_2(0)+\gamma_2(1)+\gamma_2(2) . \nonumber\\
& & 
\label{4-31b}
\eeq
\esub
The above relations lead us to 
\beq\label{4-32}
0 \leq N_0 \leq N_1 \leq N_2\ , \qquad
0\leq \gamma_2(2) \leq \gamma_2(1) \leq \gamma_2(0)\ . 
\eeq
The operators ${\wtilde \Lambda}_{\pm,0}(3)$ obey the $su(2)$-algebra and, then, the relations (\ref{4-24}) and (\ref{4-30b}) lead us to the following inequalities: 
\bsub\label{4-33}
\beq
& &0\leq \Omega-\frac{N_1+N_0}{4}\ , 
\label{4-33a}\\
& &-\left(\Omega-\frac{N_1+N_0}{4}\right) \leq -\left(\Omega-\frac{N_2-N_1}{3}-\frac{N_1+N_0}{4}\right)
\leq \Omega-\frac{N_1+N_0}{4}\ . \qquad
\label{4-33b}
\eeq
\esub
The relations (\ref{4-33a}) and (\ref{4-33b}), together with the inequality in the relation (\ref{4-32}), are rewritten as 
\bsub\label{4-34}
\beq
& &0\leq N_1 \leq 4\Omega-N_0\ , 
\label{4-34a}\\
& &0\leq N_0 \leq N_1 \leq N_2 \leq 6\Omega-\frac{1}{2}(N_1+3N_0)\ . 
\label{4-34b}
\eeq
\esub
The relation (\ref{4-32}) gives us 
\beq\label{4-35}
0\leq \gamma_2(2)\leq \gamma_2(1)\leq \gamma_2(0) \leq 2\Omega\ . 
\eeq
Needless to say, $\ket{N_0}$ and $\ket{N_1,N_0}$ are also the minimum weight states of the $su(3)$-Lipkin model. 
In the cases $(N_2=N_1=N_0)$ and $(N_2=N_1>N_0)$, $\ket{N_2,N_1,N_0}$ are reduced to $\ket{N_0}$ and $\ket{N_1,N_0}$, respectively.

\setcounter{equation}{0}
\section{The minimum weight states of the general case}

In last section, we discussed the cases of the $su(2)$- and the $su(3)$-Lipkin model. 
As are given in the relations (\ref{4-13}) and (\ref{4-25}), $\ket{N_0}$ and $\ket{N_1,N_0}$ 
are the minimum weight states of $({\wtilde \Lambda}_{\pm,0}(n))$ for $n=2$ and 3, respectively. 
The example of $\ket{N_0}$ and the explicit form of $\ket{N_1,N_0}$ are shown in the 
relation (\ref{a7}) and (\ref{4-14}), respectively. 
These two forms suggest us the following form: 
\bsub\label{5-1}\beq
\ket{N_{n-2},N_{n-3},\cdots ,N_1,N_0}
&=&
\left({\wtilde \Lambda}_+(n-1)\right)^{\frac{N_{n-2}-N_{n-3}}{n-1}}\cdot\left({\wtilde \Lambda}_+(n-2)\right)^{\frac{N_{n-3}-N_{n-4}}{n-2}}
\cdots\nonumber\\
& &\qquad\qquad\qquad\qquad \qquad
\times \left({\wtilde \Lambda}_+(2)\right)^{\frac{N_1-N_0}{2}}\ket{N_0}
\nonumber\\
&=&
\prod_{\nu=2}^{n-1}\left({\wtilde \Lambda}_+(\nu)\right)^{\frac{N_{\nu-1}-N_{\nu-2}}{\nu}}\ket{N_0}\ . \quad (n\geq 3)
\label{5-1a}
\eeq
If we adopt the form (\ref{a7}), the state (\ref{5-1a}) can be expressed as 
\beq
& &\ket{N_{n-2},N_{n-3},\cdots ,N_1,N_0}
=\prod_{\nu=1}^{n-1}\left({\wtilde \Lambda}_+(\nu)\right)^{\frac{N_{\nu-1}-N_{\nu-2}}{\nu}}\ket{N}\quad {\rm for}\quad
N_{-1}=N\ . \quad (n\geq 2)\nonumber\\
& &
\label{5-1b}
\eeq
\esub
Hereafter, we will use only the form (\ref{5-1a}). 
Therefore, our treatment is valid for $n\geq 3$. 
If the form (\ref{5-1}) is accepted, the minimum weight state of the $su(n)$-Lipkin model may be given as 
\beq\label{5-2}
& &\ket{N_{n-1},N_{n-2},N_{n-3},\cdots ,N_1,N_0}=\left({\wtilde \Lambda}_+(n)\right)^{\frac{N_{n-1}-N_{n-2}}{n}}
\ket{N_{n-2},N_{n-3},\cdots ,N_1,N_0}\ . \nonumber\\
& &
\eeq

First, let us prove the relation 
\beq\label{5-3}
{\wtilde \Lambda}_-(n)\ket{N_{n-2},N_{n-3},\cdots ,N_1,N_0}=0\ .
\eeq
For this aim, some preliminary argument is necessary. 
For the case $\nu<n$, the operator ${\tilde d}_m(n)$ introduced in the relation (\ref{3-15}) can be factorized into the form 
\beq\label{5-4}
{\tilde d}_m(n)={\tilde d}_m(\nu)\cdot {\tilde \delta}_m(n,\nu)\ , 
\eeq
\bsub\label{5-5}
\beq
& &{\tilde d}_m(\nu)={\tilde d}_{0,m}{\tilde d}_{1,m}\cdots {\tilde d}_{\nu-1,m}\ , 
\label{5-5a}\\
& &{\tilde \delta}_m(n,\nu)={\tilde d}_{\nu,m}{\tilde d}_{\nu+1,m}\cdots {\tilde d}_{n-1,m}\ . 
\label{5-5b}
\eeq
\esub
The operator $({\tilde \delta}_m^*(n,\nu),{\tilde \delta}_m(n,\nu))$ satisfies 
\beq\label{5-6}
[\ {\tilde \delta}_m^*(n,\nu)\ , \ {\tilde d}_\mu^*(\nu')\ ]=0\ , \qquad
[\ {\tilde \delta}_m(n,\nu)\ , \ {\tilde d}_\mu^*(\nu')\ ]=0 \quad {\rm for}\quad \nu'\leq \nu\ .
\eeq
The relation (\ref{5-6}) may be self-evident, because $({\tilde \delta}_m^*(n,\nu),{\tilde \delta}_m(n,\nu))$ and ${\tilde d}_\mu^*$ 
are composed from the operators different of each other. 
It can be seen in the relation (\ref{5-5}). 
The operator ${\wtilde \Lambda}_-(n)$ is expressed as 
\beq\label{5-7}
{\wtilde \Lambda}_-(n)=\sum_m{\tilde d}_m(n)=\sum_m{\tilde d}_m(\nu)\cdot{\tilde \delta}_m(n,\nu)\ . 
\eeq
Then, with the use of the relations (\ref{3-19}) and (\ref{5-6}), we have 
\bsub\label{5-8}
\beq
& &[\ {\wtilde \Lambda}_-(n)\ , \ {\wtilde \Lambda}_+(\nu)\ ]=\sum_m[\ {\tilde d}_m(\nu)\ , \ {\tilde d}_m^*(\nu)\ ]\cdot{\tilde \delta}_m(n,\nu)
\quad {\rm for}\quad \nu<n\ , 
\label{5-8a}\\
& &[\ {\tilde \delta}_m^*(n,\nu)\ , \ {\wtilde \Lambda}_+(\nu')\ ]=0\ , \qquad
[\ {\tilde \delta}_m(n,\nu)\ , \ {\wtilde \Lambda}_+(\nu')\ ]=0 \quad{\rm for}\quad \nu'\leq \nu\ . 
\label{5-8b}
\eeq
\esub
Successive use of the relation (\ref{5-8}) and the condition ${\wtilde \Lambda}_-(n)\ket{N_0}=0$ lead us 
to the relation (\ref{5-3}). 

Next, we consider that the state $\ket{N_{n-2},N_{n-3},\cdots ,N_1,N_0}$ is the eigenstate of ${\wtilde \Lambda}_0(n)$ and its eigenvalue should 
be obtained. 
The relations (\ref{3-20b}), (\ref{5-4}) and (\ref{5-6}) lead us to ${\wtilde \Lambda}_0(n)$ in the following form: 
\beq\label{5-9}
{\wtilde \Lambda}_0(n)&=&
-\frac{1}{2}\sum_m{\tilde d}_m(\nu){\tilde d}_m^*(\nu)\nonumber\\
& &+\frac{1}{2}\biggl(\sum_m{\tilde d}_m^*(\nu){\tilde d}_m(\nu)\cdot{\tilde \delta}_m^*(n,\nu){\tilde \delta}_m(n,\nu)
+{\tilde d}_m(\nu){\tilde d}_m^*(\nu)(1-{\tilde \delta}_m(n,\nu){\tilde \delta}_m^*(n,\nu))\biggl)\ , \nonumber\\ 
& &
\eeq
\bsub\label{5-10}
\beq
& &{\tilde \delta}_m^*(n,\nu){\tilde \delta}_m(n,\nu)
=({\tilde c}_{n-1,m}^* {\tilde c}_{n-1,m})\cdots({\tilde c}_{\nu,m}^*{\tilde c}_{\nu,m})\ , 
\label{5-10a}\\
& &1-{\tilde \delta}_m(n,\nu){\tilde \delta}_m^*(n,\nu)
=1-(1-{\tilde c}_{n-1,m}^* {\tilde c}_{n-1,m})\cdots(1-{\tilde c}_{\nu,m}^*{\tilde c}_{\nu,m})\ . 
\label{5-10b}
\eeq
\esub
In order to calculate $[\ {\wtilde \Lambda}_0(n)\ , \ {\wtilde \Lambda}_+(\nu)\ ]$, we must use the relation 
\beq\label{5-11}
\left[\ \sum_m{\tilde d}_m(\nu){\tilde d}_m^*(\nu)\ , \ {\wtilde \Lambda}_+(\nu)\ \right]=-{\wtilde \Lambda}_+(\nu)\ . 
\eeq
For the derivation of the relation (\ref{5-11}), we used the relations (\ref{3-18}) and (\ref{3-19}). 
With the use of the relations (\ref{5-8b}) and (\ref{5-11}), we obtain the following: 
\beq\label{5-12}
[\ {\wtilde \Lambda}_0(n)\ , \ {\wtilde \Lambda}_+(\nu)\ ]&=&
\frac{1}{2}{\wtilde \Lambda}_+(\nu)
\nonumber\\
& &+\frac{1}{2}\biggl(\sum_m[\ {\tilde d}_m^*(\nu){\tilde d}_m(\nu)\ , \ {\wtilde \Lambda}_+(\nu)\ ]\cdot{\tilde \delta}_m^*(n,\nu){\tilde \delta}_m(n,\nu)
\nonumber\\
& &\qquad
+\sum_m[\ {\tilde d}_m(\nu){\tilde d}_m^*(\nu)\ , \ {\wtilde \Lambda}_+(\nu)\ ]\cdot(1-{\tilde \delta}_m(n,\nu){\tilde \delta}_m^*(n,\nu))\biggl)\ . \qquad
\eeq
Successive use of the relation (\ref{5-12}) gives us the relation
\beq\label{5-13}
& &{\wtilde \Lambda}_0(n)\ket{N_{n-2},N_{n-3},\cdots ,N_1,N_0}=
-\lambda(n)\ket{N_{n-2},N_{n-3},\cdots ,N_1,N_0}\ , 
\nonumber\\
& &\lambda(n)=\Omega-\frac{1}{2}\left(\sum_{\nu=2}^{n-1}\frac{N_{\nu-1}-N_{\nu-2}}{\nu}+N_0\right)
\nonumber\\
& &\qquad\ 
=\Omega-\frac{1}{2}\left(\sum_{\nu=2}^{n-1}\frac{N_{\nu-1}}{\nu(\nu+1)}+\left(\frac{N_{n-2}}{n}-\frac{N_0}{2}\right)+N_0\right)\ . 
\eeq
Here, we used the relation (\ref{5-8b}) and 
\beq\label{5-14}
& &{\wtilde \Lambda}_0(n)\ket{N_0}=\frac{1}{2}(N_0-2\Omega)\ket{N_0}\ , 
\nonumber\\
& &{\tilde \delta}_m^*(n,\nu){\tilde \delta}_m(n,\nu)\ket{N_0}=0\ , \qquad
(1-{\tilde \delta}_m(n,\nu){\tilde \delta}_m^*(n,\nu))\ket{N_0}=0\ . 
\eeq
Thus, we learned that $\ket{N_{n-2},N_{n-3},\cdots ,N_1,N_0}$ is the minimum weight state of $({\wtilde \Lambda}_{\pm,0}(n))$. 
The $(N_{n-1}-N_{n-2})/n$-time operation of ${\wtilde \Lambda}_+(n)$ on this minimum weight state, we have the form (\ref{5-2}): 
\beq
& &\ket{N_{n-1},N_{n-2},N_{n-3},\cdots ,N_1,N_0}=
\left({\wtilde \Lambda}_+(n)\right)^{\frac{N_{n-1}-N_{n-2}}{n}}\ket{N_{n-2},N_{n-3},\cdots ,N_1,N_0}\nonumber\\
& &\qquad\qquad\qquad\qquad\qquad\qquad\qquad
=\prod_{\nu=2}^{n}\left({\wtilde \Lambda}_+(\nu)\right)^{\frac{N_{\nu-1}-N_{\nu-2}}{\nu}}\ket{N_0}\ , 
\label{5-15}\\
& &{\wtilde \Lambda}_0(n)\ket{N_{n-1},N_{n-2},N_{n-3},\cdots, N_1,N_0}
\nonumber\\
& & 
\qquad\qquad
=
\left(\frac{N_{n-1}-N_{n-2}}{n}-\lambda(n)\right)
\ket{N_{n-1},N_{n-2},N_{n-3},\cdots, N_1,N_0}\ . 
\label{5-16}
\eeq

Next, we will show that the state (\ref{5-15}) is the minimum weight state of the $su(n)$-Lipkin model. 
First, the following relations are derived from the relation (\ref{3-15}):
\bsub\label{5-17}
\beq
& &[\ {\wtilde S}_p(n)\ , \ {\tilde d}_{\lambda,m}^*\ ]=\delta_{\lambda p}{\tilde d}_{0,m}^*\ , \quad (p=1,\ 2,\cdots ,\ n-1)
\label{5-17a}\\
& &[\ {\wtilde S}_p^q(n)\ , \ {\tilde d}_{\lambda,m}^*\ ]=\delta_{\lambda p}{\tilde d}_{q,m}^*\ , \quad (q<p=2,\ 3,\cdots ,\ n-1)
\label{5-17b}\\
& &[\ {\wtilde S}_p^p(n)\ , \ {\tilde d}_{\lambda,m}^*\ ]=(\delta_{\lambda p}-\delta_{\lambda 0}){\tilde d}_{\lambda,m}^*\ . \quad (p=1,\ 2,\cdots ,\ n-1)
\label{5-17c}
\eeq
\esub
With the use of the relation (\ref{5-17}), we have 
\beq
& &[\ {\wtilde S}_p(n)\ , \ {\wtilde \Lambda}_+(\nu)\ ]=0\ , \qquad
[\ {\wtilde S}_p^q(n)\ , \ {\wtilde \Lambda}_+(\nu)\ ]=0 \ , 
\label{5-18}\\
& &
[\ {\wtilde S}_p^p(n)\ , \ {\wtilde \Lambda}_+(\nu)\ ]=
\left\{
\begin{array}{ll}
0 & (p\leq \nu-1) \\
-{\wtilde \Lambda}_+(\nu) & (p>\nu-1)
\end{array}\right.
\label{5-19}
\eeq
Noting the relations ${\wtilde S}_p(n)\ket{N_0}=0$, ${\wtilde S}_p^q(n)\ket{N_0}=0$ and 
${\wtilde S}_p^p(n)\ket{N_0}=-N_0\ket{N_0}$, 
we can show that the state (\ref{5-15}) is the minimum weight state of the $su(n)$-Lipkin model: 
\bsub\label{5-20}
\beq
& &{\wtilde S}_p(n)\ket{N_{n-1},N_{n-2},\cdots, N_1,N_0}=0\ , 
\label{5-20a}\\
& &{\wtilde S}_p^q(n)\ket{N_{n-1},N_{n-2},\cdots, N_1,N_0}=0\ , 
\qquad\qquad\qquad\qquad\qquad\qquad\qquad\qquad\quad
\label{5-20b}
\eeq
\esub
\beq\label{5-21}
& &{\wtilde S}_p^p(n)\ket{N_{n-1},N_{n-2},\cdots, N_1,N_0}\nonumber\\
& &\qquad\qquad
=
-\left(\sum_{\nu=1}^{p}\frac{N_{\nu-1}-N_{\nu-2}}{\nu}\right)\ket{N_{n-1},N_{n-2},\cdots, N_1,N_0}\ . \quad
(N_{-1}=0)\qquad 
\eeq
Thus, we could find the minimum weight state for the general case.

In the relations (\ref{4-20}), (\ref{4-21}), (\ref{4-34}) and (\ref{4-35}), we showed the inequalities, which the fermion numbers $N_{\nu-1}$ 
and $\gamma_{\nu-1}(p)$ in the cases of the $su(2)$- and the $su(3)$-Lipkin model should satisfy. 
As final remark of this section, we will give the inequalities for the general case. 
First, the relation between $N_{n-1}$ and $\gamma_{n-1}(p)$ for the $su(n)$-Lipkin model 
must be discussed. 
The minimum weight state $\ket{{\rm min}(n)}=\ket{N_{n-1},N_{n-2},\cdots ,N_1,N_0}$ shown in the relation (\ref{5-16}) 
gives us the following relation: 
\bsub\label{5-22}\beq
& &\gamma_{n-1}(0)=\frac{N_{n-1}-N_{n-2}}{n}+\frac{N_{n-2}-N_{n-3}}{n-1}+\cdots +\frac{N_2-N_1}{3}+\frac{N_1-N_0}{2}+N_0\ , \qquad
\label{5-22a}\\
& &\gamma_{n-1}(1)=\frac{N_{n-1}-N_{n-2}}{n}+\frac{N_{n-2}-N_{n-3}}{n-1}+\cdots +\frac{N_2-N_1}{3}+\frac{N_1-N_0}{2}\ , 
\nonumber\\
& &\qquad\qquad \vdots\nonumber\\
& &\gamma_{n-1}(n-2)=\frac{N_{n-1}-N_{n-2}}{n}+\frac{N_{n-2}-N_{n-3}}{n-1}\ , 
\nonumber\\
& &\gamma_{n-1}(n-1)=\frac{N_{n-1}-N_{n-2}}{n}\ . 
\label{5-22b}
\eeq
\esub
The relation (\ref{5-22}) is written compactly as 
\beq\label{5-23}
\gamma_{n-1}(p)=
\left\{
\begin{array}{llr}
\displaystyle \sum_{\nu=2}^n\frac{N_{\nu-1}-N_{\nu-2}}{\nu}+N_0\ , & (p=0) & \qquad\qquad\qquad \ (5.23{\rm a}) \\
\displaystyle \sum_{\nu=p+1}^n\frac{N_{\nu-1}-N_{\nu-2}}{\nu}\ , & (p=1,2,\cdots , n-1) & (5.23{\rm b})
\end{array}\right.
\nonumber
\eeq
\setcounter{equation}{23}
The relation (\ref{5-23}) is inversely expressed as 
\beq\label{5-24}
N_\nu=
\left\{
\begin{array}{llr}
\displaystyle \sum_{p=0}^\nu \gamma_{n-1}(p)-(\nu+1)\gamma_{n-1}(\nu+1)\ , & (\nu=0,1,\cdots , n-2) &\qquad\qquad (5.24{\rm a}) \\
\displaystyle \sum_{p=0}^{n-1}\gamma_{n-1}(p)\ . & (\nu=n-1) & (5.24{\rm b})
\end{array}\right.
\nonumber
\eeq
\setcounter{equation}{24}
The form (5.24b) is nothing but the relation (\ref{2-10}). 
We can rewrite (\ref{5-22}) to the following: 
\bsub\label{5-25}
\beq
& &\gamma_{n-1}(0)-\gamma_{n-1}(1)=N_0\ , \quad (p=0) 
\label{5-25a}\\
& &\gamma_{n-1}(p)-\gamma_{n-1}(p+1)=\frac{N_p-N_{p-1}}{p+1}\ , \quad
(p=1,2,\cdots ,n-2)
\label{5-25b}\\
& &\gamma_{n-1}(n-1)=\frac{N_{n-1}-N_{n-2}}{n}\ . \quad (p=n-1)
\label{5-25c}
\eeq
\esub
The right-hand side of the relation (\ref{5-25}) should be zero or positive 
and, then, we have 
\beq
& &0\leq N_0 \leq N_1 \leq \cdots \leq N_{n-2}\leq N_{n-1}\ , 
\label{5-26}\\
& &
0\leq \gamma_{n-1}(n-1) \leq \gamma_{n-1}(n-2) \leq\cdots \leq \gamma_{n-1}(1)\leq
\gamma_{n-1}(0)\ . 
\label{5-27}
\eeq
At the present, the upper limit cannot be determined.

For the determination of the upper limit, we note that $({\wtilde \Lambda}_{\pm,0}(n))$ 
obeys the $su(2)$-algebra and the relations (\ref{5-13}) and (\ref{5-16}) give 
us the following inequalities:
\beq\label{5-28}
& &\lambda(n)\geq 0\ , \nonumber\\
{\rm i.e.,}& &
\Omega-\frac{1}{2}\left(\sum_{\nu=2}^{n-1}\frac{N_{\nu-1}}{\nu(\nu+1)}
+\left(\frac{N_{n-2}}{n}-\frac{N_0}{2}\right)+N_0\right)\geq 0 \ , 
\label{5-28}
\eeq
\bsub\label{5-29}
\beq
& &-\lambda(n)\leq \frac{N_{n-1}-N_{n-2}}{n}-\lambda(n)\leq\lambda(n)\ , \qquad\qquad\ \ 
\label{5-29a}\\
{\rm i.e.,}
& &
N_{n-2}\leq N_{n-1}\leq n\left(2\Omega-\sum_{\nu=1}^{n-1}\frac{N_{\nu-1}}{\nu(\nu+1)}\right)\ . 
\label{5-29b}
\eeq
\esub
The relation (\ref{5-28}) combined with the relation (\ref{5-26}) lead us to 
\bsub\label{5-30}
\beq
& &N_0\leq 2\Omega \ , \quad (n=2)
\label{5-30a}\\
& &N_{n-2}\leq (n-1)\left(2\Omega-\sum_{\nu=1}^{n-2}\frac{N_{\nu-1}}{\nu(\nu+1)}\right)\ , 
\quad (n=3,4,\cdots)\qquad\qquad\qquad\qquad\qquad \ \ 
\label{5-30b}
\eeq
\esub
\beq\label{5-31}
0\leq N_0\leq N_1\leq \cdots \leq N_{n-1}\leq 
n\left(2\Omega-\sum_{\nu=1}^{n-1}\frac{N_{\nu-1}}{\nu(\nu+1)}\right)\ , 
\quad (n=2,3,\cdots)
\eeq
For the relations (\ref{5-29}) and (\ref{5-30}) for the cases $n=2$ and 3 reduce 
to the relations (\ref{4-20}) and (\ref{4-34}), respectively. 
The inequality (\ref{5-29a}) leads us to the following: 
\beq\label{5-32}
\gamma_{n-1}(0)\leq 2\Omega\ . 
\eeq
For the derivation, we used the relation (\ref{5-23}). 
Then, we have 
\beq\label{5-33}
0\leq\gamma_{n-1}(n-1)\leq \gamma_{n-1}(n-2)\leq \cdots \leq \gamma_{n-1}(1)
\leq \gamma_{n-1}(0)\leq 2\Omega\ . 
\eeq
Thus, we could present the minimum weight state of the general case. 
It should be noted that all relations given in this section are available for $n\geq 3$.

\setcounter{equation}{0}
\section{Discussions}

Until the present stage, we developed a possible idea how to give concrete expressions 
of the minimum weight states for the $su(n)$-Lipkin model in arbitrary 
fermion number. 
In this section, we will treat some simple examples of the minimum weight states from a viewpoint 
slightly different from that in last section. 
This argument is also in preparation for next paper (II). 
Our discussion starts in to mention that the $su(n)$-Lipkin model contains 
the $su(2)$-subalgebras. 
Its number depends on the number $n$. 
In this section, we will discuss the cases $n=2,\ 3,\ 4$ and 5. 
The case $n=2$ is the $su(2)$-algebra itself and the case $n=3$ has one $su(2)$-subalgebra. 
On the other hand, the cases $n=4$ and 5 contain two $su(2)$-algebras. 
One by one, we will show this point. 

\begin{figure}[t]
\begin{center}
\includegraphics[height=5.5cm]{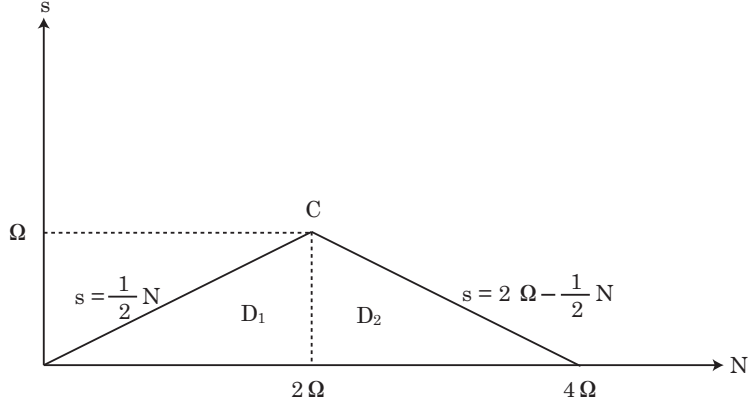}
\caption{The relation between $s$ and $N$ is shown in the inequality (\ref{6-4}).
}
\label{fig:fig2}
\end{center}
\end{figure}

In the case $n=2$, ${\wtilde S}^1(={\wtilde S}_+)$, ${\wtilde S}_1(={\wtilde S}_-)$ and ${\wtilde S}_1^1/2(={\wtilde S}_0)$ 
form the $su(2)$-algebra and ${\wtilde \Gamma}_{\rm su(2)}$ is given as 
\beq\label{6-1}
{\wtilde \Gamma}_{\rm su(2)}={\wtilde S}_+{\wtilde S}_-+{\wtilde S}_0\left({\wtilde S}_0-1\right)\ .
\eeq
The above is nothing but the original Lipkin model. 
The minimum weight state $\ket{{\rm min}(2)}$ is specified by two quantum numbers $N$ and $s$, the eigenvalues of 
${\wtilde N}$ and $-{\wtilde S}_0$: 
$\ket{{\rm min}(2)}=\ket{N;s}$. 
Of course, these two are related to the algebra. 
Then, for the orthogonal set, we have 
\beq\label{6-2}
\ket{N;ss_0}=\sqrt{\frac{(s-s_0)!}{(2s)!(s+s_0)!}}\left({\wtilde S}_+\right)^{s+s_0}\ket{N;s}\ .
\eeq
In this case, we obtain the relation 
\beq\label{6-3}
\gamma_1(0)=\frac{N}{2}+s\ (\geq 0)\ , \qquad 
\gamma_1(1)=\frac{N}{2}-s\ (\geq 0)\ . 
\eeq
Then, with the use of the inequality (\ref{4-21}), we can show that the relation (\ref{6-3}) holds in the following domains:
\bsub\label{6-4}
\beq
& &({\rm D}_1)\ \ 0\leq N \leq 2\Omega\ , \qquad 0\leq s \leq \frac{N}{2}\ , 
\label{6-4a}\\
& &({\rm D}_2)\ \ 2\Omega \leq N \leq 4\Omega\ , \qquad 0\leq s \leq 2\Omega-\frac{N}{2}\ . 
\label{6-4b}
\eeq
\esub
The above domains can be illustrated in Fig.{\ref{fig:fig2}}. 
A ``closed-shell" system appears in the case $(N=2\Omega,\ s=\Omega)$, where, in the absence of interactions, 
the level $p=0$ is occupied fully by the fermions and the level $p=1$ is vacant. 
The point C in Fig.\ref{fig:fig2} corresponds to the ``closed-shell" system. 
But, $s$ can decrease from $s=\Omega$ to $s=0$, where the level $p=0$ 
and $p=1$ are occupied in equal fermion number $\Omega$. 
  
%
%

\begin{figure}[t]
\begin{center}
\includegraphics[height=5.5cm]{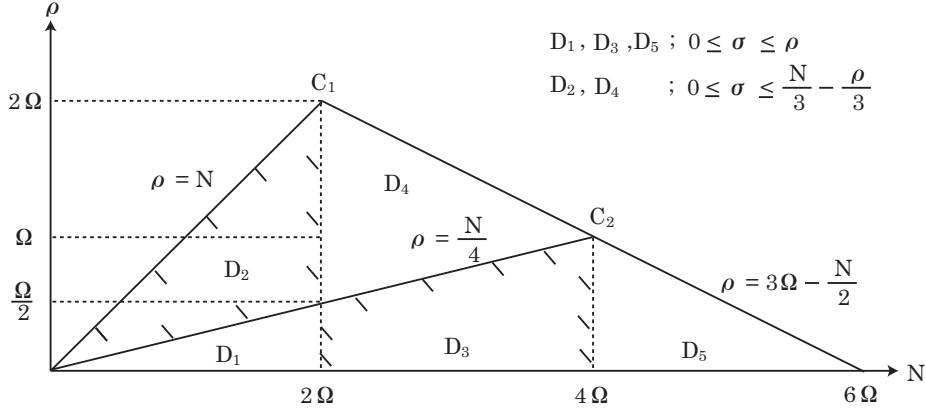}
\caption{The domains depicted in (\ref{6-9}) are illustrated. 
}
\label{fig:fig3}
\end{center}
\end{figure}

Next, we treat the case $n=3$. 
The operators ${\wtilde S}_1^2 (={\wtilde S}_+)$, ${\wtilde S}_2^1 (={\wtilde S}_-)$ 
and $({\wtilde S}_2^2-{\wtilde S}_1^1)/2 (={\wtilde S}_0)$ form the $su(2)$-subalgebra and, further, 
we have the scalar ${\wtilde R}_0$ with respect to $({\wtilde S}_{\pm,0})$ in the form 
\beq\label{6-5}
{\wtilde R}_0=\frac{1}{2}\left({\wtilde S}_2^2+{\wtilde S}_1^1\right)\ . \qquad
\left(\ \left[\ {\wtilde S}_{\pm,0}\ , \ {\wtilde R}_0\ \right]=0\ \right)
\eeq
The Casimir operator ${\wtilde \Gamma}_{\rm su(3)}$ is expressed as 
\beq\label{6-6}
{\wtilde \Gamma}_{\rm su(3)}=
\left({\wtilde S}^2{\wtilde S}_2+{\wtilde S}^1{\wtilde S}_1\right)
+\left({\wtilde S}_+{\wtilde S}_-+{\wtilde S}_0\left({\wtilde S}_0-1\right)\right)
+\frac{1}{3}{\wtilde R}_0\left({\wtilde R}_0-3\right)\ . 
\eeq
In addition to $N$, $\ket{{\rm min}(3)}$ can be specified by the eigenvalues of 
${\wtilde S}_0$ and ${\wtilde R}_0$, $-\sigma$ and $-\rho$, respectively: 
$\ket{{\rm min}(3)}=\ket{N;\rho,\sigma}$. 
Then, we have 
\beq\label{6-7}
\ket{N;\rho,\sigma\sigma_0}
=\sqrt{\frac{(\sigma-\sigma_0)!}{(2\sigma)!(\sigma+\sigma_0)!}}
\left({\wtilde S}_+\right)^{\sigma+\sigma_0}\ket{N;\rho,\sigma}\ . 
\eeq
Therefore, for constructing the orthogonal sets, we, further, must take account of 
$({\wtilde S}^2, {\wtilde S}^1)$. 
It will be discussed in (II). 
In the present, we have the relation
\beq\label{6-8}
\gamma_2(0)=\frac{N}{3}+\frac{2\rho}{3}\ , \qquad
\gamma_2(1)=\frac{N}{3}-\frac{\rho}{3}+\sigma\ , \qquad
\gamma_2(2)=\frac{N}{3}-\frac{\rho}{3}-\sigma .
\eeq
The inequality (\ref{4-32}) leads us to the following domains for the relation (\ref{6-8}). 
\bsub\label{6-9}
\beq
& &{\rm (i)}\ \ \ 0\leq N \leq 2\Omega\nonumber\\
& &\qquad ({\rm D}_1)\ \ 0\leq \rho \leq \frac{N}{4}\ , \ \ 0\leq \sigma \leq \rho\ , 
\quad ({\rm D}_2)\ \ \frac{N}{4}\leq \rho \leq N\ , \ \ 
0\leq \sigma \leq \frac{N}{3}-\frac{\rho}{3}\ , 
\label{6-9a}\\
& &{\rm (ii)}\ \ 2\Omega\leq N \leq 4\Omega\nonumber\\
& &\qquad ({\rm D}_3)\ \ 0\leq \rho \leq \frac{N}{4}\ , \ \ 0\leq \sigma \leq \rho \ ,
\quad ({\rm D}_4)\ \ \frac{N}{4}\leq \rho \leq 3\Omega-\frac{N}{2}\ , \ \ 
0\leq \sigma \leq \frac{N}{3}-\frac{\rho}{3}\ , \nonumber\\
& &
\label{6-9b}\\
& &{\rm (iii)}\ \ 4\Omega\leq N \leq 6\Omega\nonumber\\
& &\qquad ({\rm D}_5)\ \ 0\leq \rho \leq 3\Omega-\frac{N}{2}\ , \ \ 0\leq \sigma \leq \rho 
\ . 
\label{6-9c}
\eeq
\esub
The above domains are illustrated in Fig.\ref{fig:fig3}. 
The present case contains two ``closed-shell" systems. 
First appears at the point C$_1$ in Fig.\ref{fig:fig3} $(N=2\Omega,\ \rho=2\Omega,\ \sigma=0)$. 
Only the level $p=0$ is occupied. 
Second appears at the point C$_2$ in Fig.\ref{fig:fig3} 
($N=4\Omega,\ \rho=\Omega,\ \sigma=\Omega)$. 
In this case, the levels $p=0$ and 1 are occupied fully. 
However, by changing the values of $\rho$ and $\sigma$, we can produce various fermion number distributions.

%
%

\begin{figure}[t]
\begin{center}
\includegraphics[height=7cm]{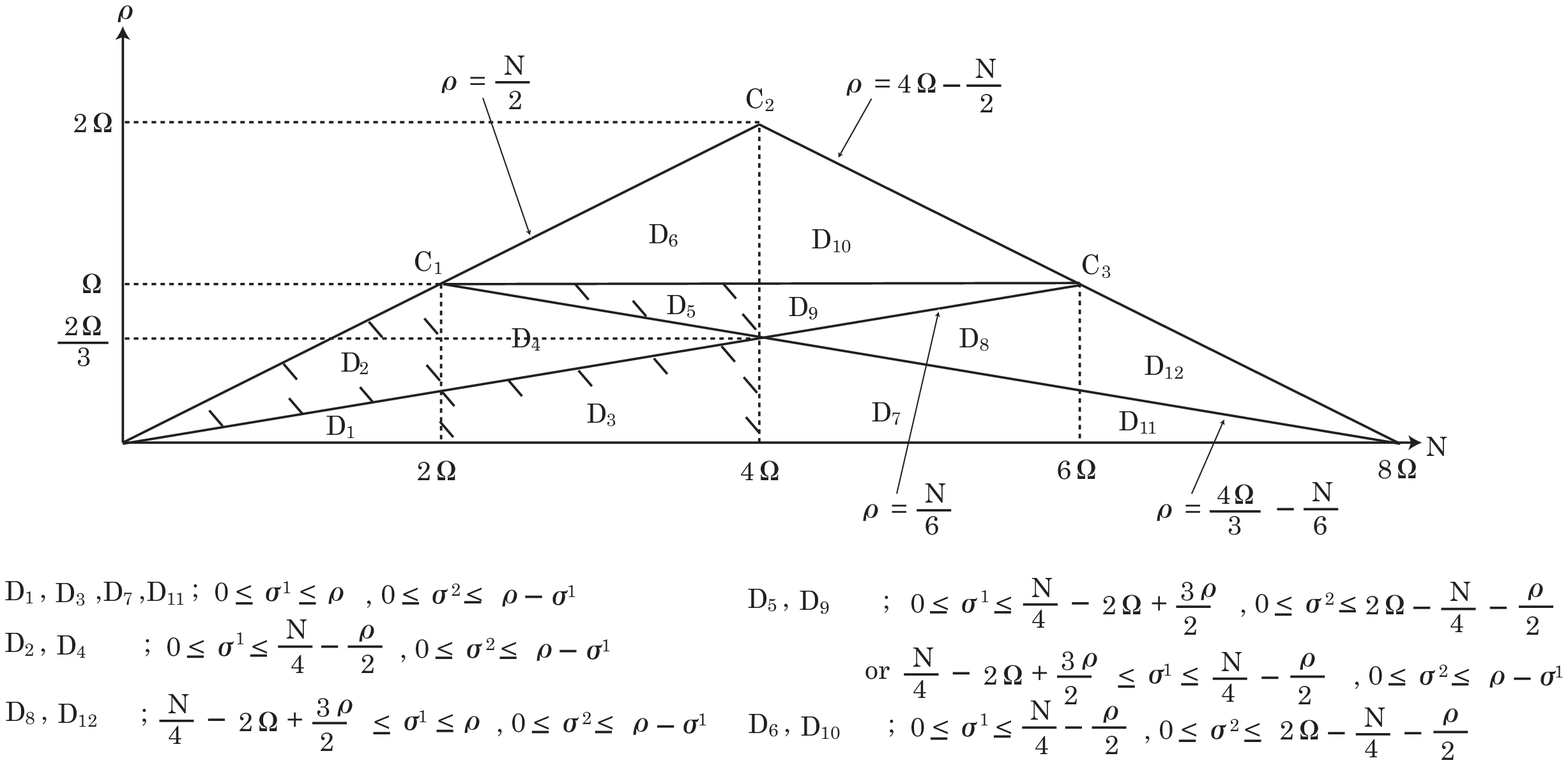}
\caption{The domains depicted in (\ref{6-15}) are illustrated. 
}
\label{fig:fig4}
\end{center}
\end{figure}

Third is concerned with the case $n=4$, where there exist two $su(2)$-subalgebras: 
${\wtilde S}_2^3 (={\wtilde S}_+(1))$, ${\wtilde S}_3^2 (={\wtilde S}_-(1))$, $({\wtilde S}_3^3-{\wtilde S}_2^2)/2 (={\wtilde S}_0(1))$ and 
${\wtilde S}^1 (={\wtilde S}_+(2))$, ${\wtilde S}_1 (={\wtilde S}_-(2))$, ${\wtilde S}_1^1/2 (={\wtilde S}_0(2))$. 
Further, we denote the addition of the above two as 
\beq\label{6-10}
{\wtilde S}_{\pm,0}={\wtilde S}_{\pm,0}(1)+{\wtilde S}_{\pm,0}(2)\ .
\eeq
This case gives us one scalar with respect to $({\wtilde S}_{\pm,0})$: 
\beq\label{6-11}
{\wtilde R}_0=\frac{1}{2}\left({\wtilde S}_3^3+{\wtilde S}_2^2-{\wtilde S}_1^1\right)\ . 
\eeq
The Casimir operator ${\wtilde \Gamma}_{\rm su(4)}$ is written as 
\beq\label{6-12}
{\wtilde \Gamma}_{\rm su(4)}&=&
\left({\wtilde S}^3{\wtilde S}_3+{\wtilde S}^2{\wtilde S}_2+{\wtilde S}_1^3{\wtilde S}_3^1+{\wtilde S}_1^2{\wtilde S}_2^1\right)
\nonumber\\
& &+\sum_{i=1,2}\left({\wtilde S}_+(i){\wtilde S}_-(i)+{\wtilde S}_0(i)\left({\wtilde S}_0(i)-1\right)\right)
+\frac{1}{2}{\wtilde R}_0\left({\wtilde R}_0-4\right)\ . 
\eeq
The minimum weight state $\ket{{\rm min}(4)}$ can be expressed as $\ket{N;\rho,\sigma^1,\sigma^2}$. 
Here, of course, $\rho$, $\sigma^1$ and $\sigma^2$ denote the eigenvalues of $-{\wtilde R}_0$, 
$-{\wtilde S}_0(1)$ and $-{\wtilde S}_0(2)$, respectively. 
Then, we have the following state: 
\beq\label{6-13}
\ket{N;\rho,\sigma^1 ,\sigma^2, \sigma\sigma_0}
&=&
\sum_{\sigma_0^1,\sigma_0^2}
\langle \sigma^1\sigma_0^1,\sigma^2\sigma_0^2 \ket{\sigma\sigma_0}
\sqrt{\frac{(\sigma^1-\sigma_0^1)!}{(2\sigma^1)!(\sigma^1+\sigma_0^1)!}}
\sqrt{\frac{(\sigma^2-\sigma_0^2)!}{(2\sigma^2)!(\sigma^2+\sigma_0^2)!}}\nonumber\\
& &\times\left({\wtilde S}_+(1)\right)^{\sigma^1+\sigma_0^1}\left({\wtilde S}_+(2)\right)^{\sigma^2+\sigma_0^2}
\ket{N;\rho,\sigma^1,\sigma^2}\ . 
\eeq
Then, the role of ${\wtilde S}^3$, ${\wtilde S}^2$, ${\wtilde S}_1^3$ and ${\wtilde S}_1^2$ becomes interesting for constructing the orthogonal sets. 
In the present case, we can derive the relation
\beq\label{6-14}
& &\gamma_4(0)=\frac{N}{4}+\frac{\rho}{2}+\sigma^2\ , \qquad
\gamma_4(1)=\frac{N}{4}+\frac{\rho}{2}-\sigma^2\ , \nonumber\\
& &\gamma_4(2)=\frac{N}{4}-\frac{\rho}{2}+\sigma^1\ , \qquad
\gamma_4(3)=\frac{N}{4}-\frac{\rho}{2}-\sigma^1\ . 
\eeq
The inequality (\ref{5-33}) gives the following 12 domains:
\bsub\label{6-15}
\beq
& &{\rm (i)}\ \ \ 0\leq N \leq 2\Omega\nonumber\\
& &\qquad ({\rm D}_1)\ \ 0\leq \rho \leq \frac{N}{6}\ ,  
\qquad ({\rm D}_2)\ \ \frac{N}{6}\leq \rho \leq \frac{N}{2}\ ,  
\label{6-15a}\\
& &{\rm (ii)}\ \ 2\Omega\leq N \leq 4\Omega\nonumber\\
& &\qquad ({\rm D}_3)\ \ 0\leq \rho \leq \frac{N}{6}\ ,  
\qquad ({\rm D}_4)\ \ \frac{N}{6}\leq \rho \leq \frac{4\Omega}{3}-\frac{N}{6}\ , \nonumber\\
& &\qquad ({\rm D}_5)\ \ \frac{4\Omega}{3}-\frac{N}{6} \leq \rho \leq \Omega\ ,  
\qquad ({\rm D}_6)\ \ \Omega \leq \rho \leq \frac{N}{2}\ , 
\label{6-15b}\\
& &{\rm (iii)}\ \ 4\Omega\leq N \leq 6\Omega\nonumber\\
& &\qquad ({\rm D}_7)\ \ 0\leq \rho \leq \frac{4\Omega}{3}-\frac{N}{6}\ , 
\qquad ({\rm D}_8)\ \ \frac{4\Omega}{3}-\frac{N}{6} \leq \rho \leq \frac{N}{6}\ , \nonumber\\
& &\qquad ({\rm D}_9)\ \ \frac{N}{6} \leq \rho \leq \Omega\ , 
\qquad ({\rm D}_{10})\ \ \Omega \leq \rho \leq 4\Omega-\frac{N}{2}\ , 
\label{6-15c}\\
& &{\rm (iv)}\ \ 6\Omega\leq N \leq 8\Omega\nonumber\\
& &\qquad ({\rm D}_{11})\ \ 0\leq \rho \leq 4\Omega-\frac{N}{2}\ ,  
\qquad ({\rm D}_{12})\ \ \frac{4\Omega}{3}-\frac{N}{6}\leq \rho \leq 4\Omega-\frac{N}{2}\ .  
\label{6-15d}
\eeq
\esub
The above is illustrated in Fig.{\ref{fig:fig4}}. 
Composed with the above, it is complicated. 
%
%
Three ``closed-shell" systems appear in this case: 
C$_1$, $N=2\Omega$, $\rho=\Omega$, $\sigma^1=0$, $\sigma^2=0$, 
C$_2$, $N=4\Omega$, $\rho=2\Omega$, $\sigma^1=0$, $\sigma^2=0$ and C$_3$, $N=6\Omega$, $\rho=\Omega$, $\sigma^1=\Omega$, $\sigma^2=0$. 
By changing the values of $\rho$, $\sigma^1$ and $\sigma^2$, we can produce various fermion number distribution.

Finally, we will treat the case $n=5$. 
In this case, we have also two $su(2)$-subalgebras: 
${\wtilde S}_3^4 (={\wtilde S}_+(1))$, ${\wtilde S}_4^3 (={\wtilde S}_-(1))$, 
$({\wtilde S}_4^4-{\wtilde S}_3^3)/2 (={\wtilde S}_0(1))$ and ${\wtilde S}_1^2 (={\wtilde S}_+(2))$, ${\wtilde S}_2^1 (={\wtilde S}_-(2))$, 
$({\wtilde S}_2^2-{\wtilde S}_1^1)/2 (={\wtilde S}_0(2))$. 
For these two, we also use ${\wtilde S}_{\pm,0}$ given in the relation (\ref{6-10}). 
However, the present case contains two scalars with respect to $({\wtilde S}_{\pm,0})$: 
\beq\label{6-17}
{\wtilde R}_0(1)=\frac{1}{2}\left({\wtilde S}_4^4+{\wtilde S}_3^3-{\wtilde S}_2^2-{\wtilde S}_1^1\right)\ , \qquad
{\wtilde R}_0(2)=\frac{1}{2}\left({\wtilde S}_4^4+{\wtilde S}_3^3+{\wtilde S}_2^2+{\wtilde S}_1^1\right) \ . 
\eeq
The Casimir operator ${\wtilde \Gamma}_{\rm su(5)}$ can be expressed as 
\beq\label{6-18}
{\wtilde \Gamma}_{\rm su(5)}
&=&
\left({\wtilde S}^4{\wtilde S}_4+{\wtilde S}^3{\wtilde S}_3+{\wtilde S}^2{\wtilde S}_2+{\wtilde S}^1{\wtilde S}_1
+{\wtilde S}_1^4{\wtilde S}_4^1+{\wtilde S}_1^3{\wtilde S}_3^1+{\wtilde S}_2^4{\wtilde S}_4^2+{\wtilde S}_2^3{\wtilde S}_3^2\right)
\nonumber\\
& &+\sum_{i=1,2}\left({\wtilde S}_+(i){\wtilde S}_-(i)+{\wtilde S}_0(i)\left({\wtilde S}_0(i)-1\right)\right)
\nonumber\\
& &+\frac{1}{2}{\wtilde R}_0(1)\left({\wtilde R}_0(1)-4\right)
+\frac{1}{10}{\wtilde R}_0(2)\left({\wtilde R}_0(2)-10\right)\ . 
\eeq
The minimum weight state is specified by $\rho^1$, $\rho^2$, $\sigma^1$ and $\sigma^2$, which are the eigenvalues of 
$-{\wtilde R}_0(1)$, $-{\wtilde R}_0(2)$, $-{\wtilde S}_0(1)$ and $-{\wtilde S}_0(2)$, respectively: 
$\ket{{\rm min}(5)}=\ket{N;\rho^1,\rho^2,\sigma^1,\sigma^2}$. 
Then, we have 
\beq\label{6-19}
\ket{N;\rho^1,\rho^2,\sigma^1,\sigma^2, \sigma \sigma_0}
&=&
\sum_{\sigma_0^1,\sigma_0^2}\langle \sigma^1\sigma_0^1\sigma^2\sigma_0^2\ket{\sigma\sigma_0}
\sqrt{\frac{(\sigma^1-\sigma_0^1)!}{(2\sigma^1)!(\sigma^1+\sigma_0^1)!}}
\sqrt{\frac{(\sigma^2-\sigma_0^2)!}{(2\sigma^2)!(\sigma^2+\sigma_0^2)!}}
\nonumber\\
& &\times \left({\wtilde S}_+(1)\right)^{\sigma^1+\sigma_0^1} 
\left({\wtilde S}_+(2)\right)^{\sigma^2+\sigma_0^2}\ket{N;\rho^1,\rho^2,\sigma^1,\sigma^2}\ . 
\eeq
Of course, the role of ${\wtilde S}^4$, ${\wtilde S}^3$, ${\wtilde S}^2$, ${\wtilde S}^1$, ${\wtilde S}_1^4$, ${\wtilde S}_1^3$, ${\wtilde S}_2^4$ and 
${\wtilde S}_2^3$ 
must be investigated.
In the case $n=5$, we have the following relation: 
\beq\label{6-20}
& &\gamma_5(0)=\frac{N}{5}+\frac{2}{5}\rho^2\ , 
\nonumber\\
& &\gamma_5(1)=\frac{N}{5}+\frac{1}{2}\rho^1-\frac{1}{10}\rho^2+\sigma^2\ , \qquad
\gamma_5(2)=\frac{N}{5}+\frac{1}{2}\rho^1-\frac{1}{10}\rho^2-\sigma^2\ ,
\nonumber\\
& &\gamma_5(3)=\frac{N}{5}-\frac{1}{2}\rho^1-\frac{1}{10}\rho^2+\sigma^1\ , 
\qquad
\gamma_5(4)=\frac{N}{5}-\frac{1}{2}\rho^1-\frac{1}{10}\rho^2-\sigma^1\ .  
\eeq
In this case, we also use the inequality (\ref{5-33}). 
But, different from the case $n=4$, we cannot give the relations between $N$ and $\rho^1$ also between $N$ and $\rho^2$, respectively. 
We give the relation between $\rho^1$ and $\rho^2$ by regarding $N$ as a parameter. 
Inequality (\ref{5-33}) except for $\gamma_5(0) \leq 2\Omega$ leads us to the following:  
\bsub\label{6-21}
\beq
& &({\rm D}_{\rm I})\quad \ \rho^1\geq \frac{2N}{15}-\frac{\rho^2}{15}\ , \qquad
\rho^1\geq \frac{\rho^2}{3}\ , \qquad
\rho^1\geq \frac{N}{10}+\frac{\rho^2}{5}\ , 
\label{6-21a}\\
& &({\rm D}_{\rm II})\quad \ \rho^1\geq \frac{2N}{15}-\frac{\rho^2}{15}\ , \qquad
\rho^1\geq \frac{\rho^2}{3}\ , \qquad
\rho^1\leq \frac{N}{10}+\frac{\rho^2}{5}\ , 
\label{6-21b}\\
& &({\rm D}_{\rm III})\quad \ \rho^1\geq \frac{2N}{15}-\frac{\rho^2}{15}\ , \qquad
\rho^1\leq \frac{\rho^2}{3}\ , 
\label{6-21c}\\
& &({\rm D}_{\rm IV})\quad \ \rho^1\leq \frac{2N}{15}-\frac{\rho^2}{15}\ , \qquad
\rho^1\geq \frac{\rho^2}{3}\ , 
\label{6-21d}\\
& &({\rm D}_{{\rm V}})\quad \ \rho^1\leq \frac{2N}{15}-\frac{\rho^2}{15}\ , \qquad
\rho^1\leq \frac{\rho^2}{3}\ . 
\label{6-21e}
\eeq
\esub
%
\begin{figure}[t]
\begin{center}
\includegraphics[height=7.5cm]{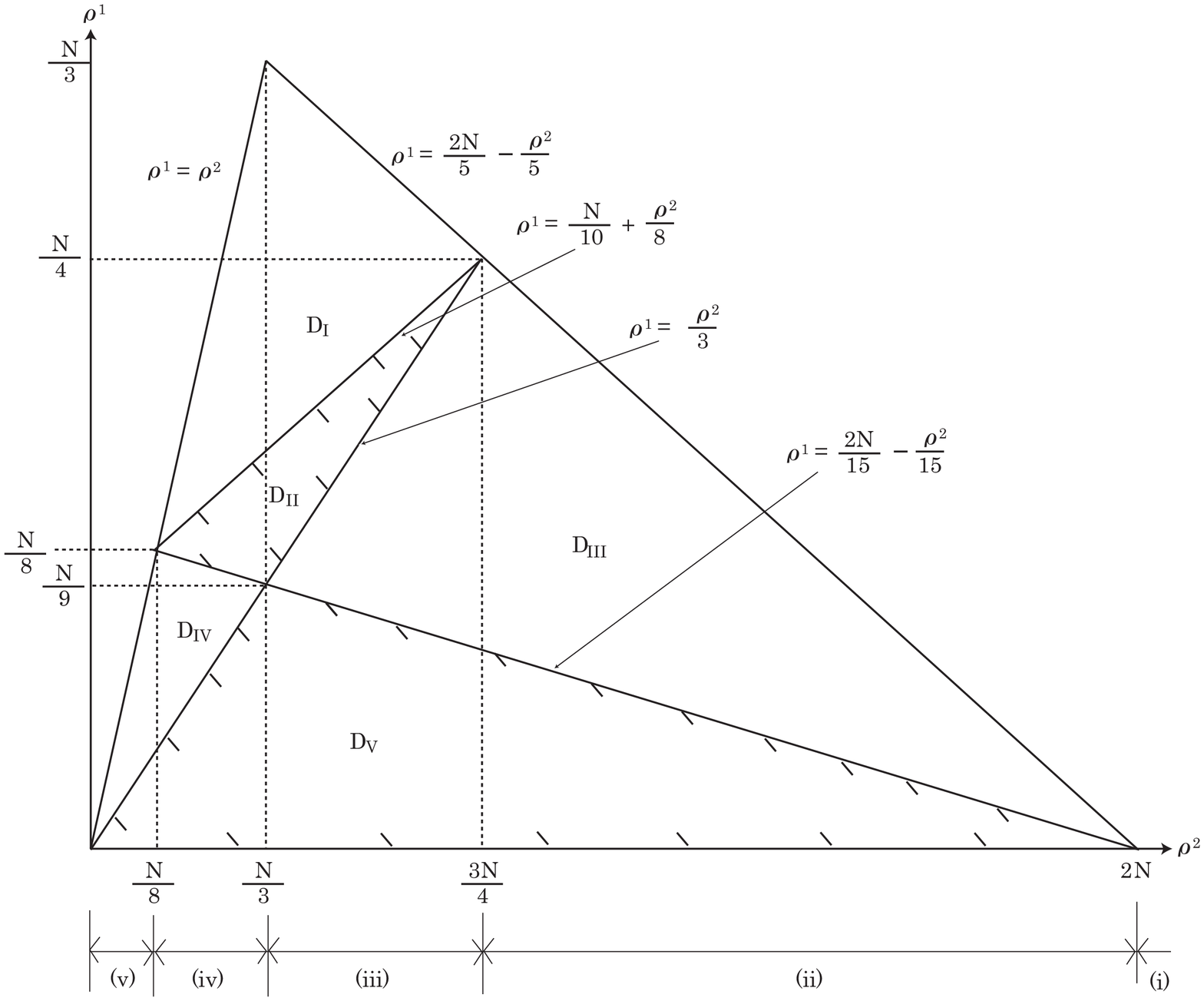}
\caption{The domains depicted in (\ref{6-21}) are illustrated. 
}
\label{fig:fig5}
\end{center}
\end{figure}
%
The relation (\ref{6-21}) is illustrated in Fig.\ref{fig:fig5}. 
In each domain, $\sigma^1$ and $\sigma^2$ obey the inequality 
\bsub\label{6-22}
\beq
& &({\rm D}_{\rm I})\quad \ 0\leq \sigma^1 \leq \frac{N}{5}-\frac{\rho^2}{10}-\frac{\rho^1}{2}\ , \qquad
0\leq \sigma^2 \leq \frac{\rho^2}{2}-\frac{\rho^1}{2}\ , 
\label{6-22a}\\
& &({\rm D}_{\rm II})\quad \ 0\leq \sigma^1 \leq \frac{3\rho^1}{2}-\frac{\rho^2}{2}\ , \qquad
0\leq \sigma^2 \leq \frac{\rho^2}{2}-\frac{\rho^1}{2}\ , \nonumber\\
& &\quad {\rm or}\qquad
\frac{3\rho^1}{2}-\frac{\rho^2}{2} \leq \sigma^1 \leq \frac{N}{5}-\frac{\rho^2}{10}-\frac{\rho^1}{2}\ , \qquad
0\leq \sigma^2 \leq \rho^1-\sigma^1\ , 
\label{6-22b}\\
& &({\rm D}_{\rm III})\quad \ 0\leq \sigma^1 \leq \frac{N}{5}-\frac{\rho^2}{10}-\frac{\rho^1}{2}\ , \qquad
0\leq \sigma^2 \leq \rho^1-\sigma^1\ , 
\label{6-22c}\\
& &({\rm D}_{\rm IV})\quad \ 0\leq \sigma^1 \leq \frac{3\rho^1}{2}-\frac{\rho^2}{2}\ , \qquad
0\leq \sigma^2 \leq \frac{\rho^2}{2}-\frac{\rho^1}{2}\ , \nonumber\\
& & \quad {\rm or}\qquad
\frac{3\rho^1}{2}-\frac{\rho^2}{2} \leq \sigma^1 \leq \rho^1\ , \qquad
0\leq \sigma^2 \leq \rho^1-\sigma^1\ , 
\label{6-22d}\\
& &({\rm D}_{{\rm V}})\quad \ 0  \leq \sigma^1 \leq \rho^1\ , \qquad
0 \leq \sigma^2 \leq \rho^1-\sigma^1\ . 
\label{6-22e}
\eeq
\esub
Inequality $\gamma_5(0)\leq 2\Omega$ gives us the relation 
\beq\label{6-23}
\rho^2 \leq 5\Omega-\frac{N}{2}\ (=\rho)\ .
\eeq
The relation (\ref{6-23}) does not depend on $\rho^1$, $\sigma^1$ and $\sigma^2$.

%
\begin{figure}[t]
\begin{center}
\includegraphics[height=8.5cm]{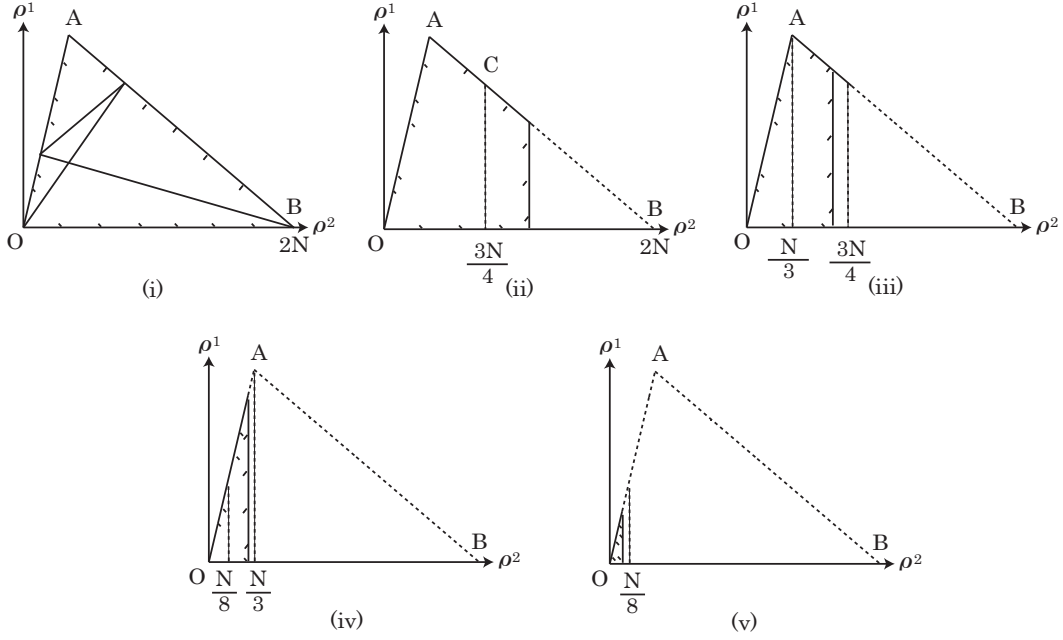}
\caption{It is shown that the relation (\ref{6-22}) is applied to the domains surrounded by short oblique lines. 
}
\label{fig:fig6}
\end{center}
\end{figure}
%

Combining $\rho$ defined in the relation (\ref{6-23}) with the regions (i) $\sim$ (v) in the $\rho^2$-axis of Fig.\ref{fig:fig5}, we have
\bsub\label{6-24}
\beq
& &
({\rm v})\ \ 0\leq \rho \leq \frac{N}{8}\ , \quad
({\rm iv})\ \ \frac{N}{8}\leq \rho \leq \frac{N}{3}\ , \quad
({\rm iii})\ \ \frac{N}{3}\leq \rho \leq \frac{3N}{4}\ , \nonumber\\
& &
({\rm ii})\ \ \frac{3N}{4}\leq \rho \leq 2N\ , \quad
({\rm i})\ \ 2N\leq \rho \ .
\label{6-24a}
\eeq
The relation (\ref{6-24a}) is reduced to 
\beq
& &
({\rm i})\ \ 0\leq N \leq 2\Omega \ , \quad
({\rm ii})\ \ 2\Omega \leq N \leq 4\Omega \ , \quad
({\rm iii})\ \ 4\Omega \leq N \leq 6\Omega \ , \nonumber\\
& &
({\rm iv})\ \ 6\Omega \leq N \leq 8\Omega \ , \quad
({\rm v})\ \ 8\Omega \leq N \leq 10\Omega \ .
\label{6-24b}
\eeq
\esub
The relation (\ref{6-24b}) is arranged in the inverted order. 
It does not necessarily follow that each region covers the whole domains shown in Fig.\ref{fig:fig5} ($\triangle$OAB in Fig.\ref{fig:fig6}).
We show this feature in Fig.\ref{fig:fig6}.

The relation (\ref{6-22}) should be applied to the domains surrounded by short oblique lines in Fig.\ref{fig:fig6}. 
Four ``closed-shell" systems appear in the present case; 
C$_1$($N=2\Omega,\ \rho^2=4\Omega,\ \rho^1=0,\ \sigma^2=\sigma^1=0)$, C$_2$($N=4\Omega,\ \rho^2=3\Omega,\ \rho^1=\Omega,\ 
\sigma^2=\Omega,\ \sigma^1=0$), C$_3$($N=6\Omega,\ \rho^2=2\Omega,\ \rho^1=0,\ \sigma^2=\sigma^1=0$) and 
C$_4$($N=8\Omega,\ \rho^2=\Omega,\ \rho^1=\Omega,\ \sigma^2=0,\ \sigma^1=\Omega$). 
By changing the values of $\rho^2$, $\rho^1$, $\sigma^2$ and $\sigma^1$, we can produce various fermion number distribution.

In this section, we have presented the structure of the minimum weight states for the cases $n=2\sim 5$ in the form slightly different from that given in {\bf 5}. 
The basic idea comes from the introduction of the $su(2)$-subalgebras and the scalar operators 
defined in the relations (\ref{6-5}), (\ref{6-19}) and (\ref{6-17}). 
In (II), we will discuss the cases with arbitrary values of $n$. 
Of course, the scalar operators are generalized.

\section*{Acknowledgment}


Two of the authors (Y.T. and M.Y.) would like to express their thanks to 
Professor J. da Provid\^encia and Professor C. Provid\^encia, two of co-authors of this paper, 
for their warm hospitality during their visit to Coimbra in spring of 2015. 
The author, M.Y., would like to express his sincere thanks to Mrs K. Yoda-Ono for her cordial encouragement. 
The authors, Y.T., is partially supported by the Grants-in-Aid of the Scientific Research 
(No.26400277) from the Ministry of Education, Culture, Sports, Science and 
Technology in Japan.

\appendix
\section{A possible example of the state $\ket{N_0}$ introduced in the relation (\ref{4-1})}

In this Appendix, the state $\ket{N_0}$ is presented through the eigenvalue problem of 
$({\wtilde \Lambda}_{\pm,0}(1))$ defined in the relation (\ref{3-12}). 
The level $p=0$ consists of $2\Omega$ single-particle states $m=-j,\ -j+1,\cdots ,\ j-1,\ j\ (2\Omega=2j+1)$. 
These states can be divided into two groups. 
One consists of ($m_1, \ m_2,\cdots ,\ m_{\Omega})$ and the other 
$({\bar m}_1,\ {\bar m}_2,\cdots ,\ {\bar m}_{\Omega})$. 
We regard the state ${\bar m}_i$ as the partner of $m_i\ (i=1,\ 2,\cdots ,\ \Omega)$. 
The choice is arbitrary and, for example, all of $m_i$ and ${\bar m}_i$ are positive and negative, respectively. 
Under the above classification, we define the state 
\beq\label{a1}
\kket{m_{v}, m_{v-1},\cdots ,m_2, m_1}
={\tilde c}_{m_v}^*{\tilde c}_{m_{v-1}}^*\cdots {\tilde c}_{m_2}^*{\tilde c}_{m_1}^*\ket{0}\ . 
\eeq 
Here, the index $p=0$ was omitted and we fix the ordering of $m_i$ appropriately, for example, $m_v>m_{v-1}>\cdots >m_2 >m_1$. 
It may be self-evident that the state 
(\ref{a1}) is not the minimum weight state of $({\wtilde \Lambda}_{\pm,0}(1))$. 
Then, by replacing ${\tilde c}_{m}^*$ with ${\wtilde D}_m^*$, we introduce the following state:
\beq
& &\ket{m_v,m_{v-1},\cdots ,m_2,m_1}={\wtilde D}_{m_v}^*{\wtilde D}_{m_{v-1}}^*\cdots {\wtilde D}_{m_2}^*{\wtilde D}_{m_1}^*\ket{0} \ , 
\label{a2}\\
& &{\wtilde D}_m^*=\frac{1}{\sqrt{2}}({\tilde d}_{m}^*-{\tilde d}_{\bar m}^*)
=\frac{1}{\sqrt{2}}(e_m{\tilde c}_m^*-e_{\bar m}{\tilde c}_{\bar m}^*)\ . 
\label{a3}
\eeq
The operator ${\wtilde D}_m^*$ satisfies 
\beq\label{a4}
[\ {\wtilde \Lambda}_-(1)\ , \ {\wtilde D}_m^*\ ]=-\sqrt{2}({\tilde c}_m^*{\tilde c}_m-{\tilde c}_{\bar m}^*{\tilde c}_{\bar m})\ . 
\eeq
Therefore, we have 
\beq
& &{\wtilde \Lambda}_-(1)\ket{m_v,m_{v-1},\cdots , m_2,m_1}=0\ , 
\label{a5}\\
& &0\leq v \leq \Omega\ . 
\label{a6}
\eeq
Next, we consider the state 
\beq\label{a7}
\ket{N_0;m_v,m_{v-1},\cdots ,m_2,m_1}=\left({\wtilde \Lambda}_+(1)\right)^{N_0-v}\ket{m_v,m_{v-1},\cdots , m_2,m_1}\ . 
\eeq
The relation (\ref{a7}) satisfies 
\beq\label{a8}
{\wtilde \Lambda}_0(1)\ket{N_0;m_v,m_{v-1},\cdots ,m_2,m_1}=-(\Omega-N_0)\ket{N_0;m_v,m_{v-1},\cdots ,m_2,m_1}\ , 
\eeq
i.e., 
\beq\label{a9}
{\wtilde N}(1)\ket{N_0;m_v,m_{v-1},\cdots ,m_2,m_1}=N_0\ket{N_0;m_v,m_{v-1},\cdots ,m_2,m_1}\ . 
\eeq

The above is nothing but the relation (\ref{4-1}). 
The present eigenvalue problem gives us 
\beq\label{a10}
-(\Omega-v) \leq N_0-\Omega \leq \Omega-v\ . 
\eeq 
Combining with the inequality (\ref{a6}), we have the inequality 
\beq\label{a11}
0 \leq v \leq \Omega\ , \qquad v \leq N_0 \leq 2\Omega-v\ . 
\eeq
The above is a possible example of $\ket{N_0}$.

\end{document}